\documentclass[aps,prl,twocolumn,amsmath,amssymb,longbibliography,superscriptaddress,floatfix]{revtex4-2}
\usepackage{amsmath,amssymb,amsfonts,bm}
\usepackage{ifthen}
\usepackage{graphicx}
\usepackage[colorlinks=true,breaklinks=true]{hyperref}
\usepackage{mathtools}
\usepackage{braket}
\usepackage{mathrsfs}
\usepackage{xcolor}
\usepackage{xparse}
\usepackage{orcidlink}
\usepackage{soul}
\usepackage{cancel}

\NewDocumentCommand{\s}{o}{%
  \IfNoValueTF{#1}
    {\overline{S}}
    {\overline{S(#1)}}
}
\newcommand{\be}{\begin{equation}}
\newcommand{\ee}{\end{equation}}
\newcommand{\average}[1]{\left\langle #1 \right\rangle}
\newcommand{\overbar}[1]{\overline{#1}}

\def\t{\mathsf{t}}
\def\f{\mathsf {f}}
\def\z{\mathrm{z}}
\newcommand{\xvec}{\mathbf{x}}
\newcommand{\yvec}{\mathbf{y}}
\newcommand{\tyvec}{\tilde{\yvec}}
\def\zz{\mathsf{z}}
\def\n{n_s}
\def\c{\mathsf{c}}
\newcommand{\II}{I\!I}

\begin{document}

\title{Bayesian Monitoring of a Diffusive Particle in One Dimension}

\author{Federico Gerbino~\orcidlink{0009-0008-1485-3764}}
\email{federico.gerbino@universite-paris-saclay.fr}
\affiliation{Laboratoire de Physique Théorique et Modèles Statistiques, Université Paris-Saclay, CNRS, 91405 Orsay, France}

\author{Guido Giachetti~\orcidlink{0000-0002-4928-7693}}
\affiliation{Laboratoire de Physique de l'\'Ecole Normale Sup\'erieure, CNRS, ENS $\&$ PSL University, Sorbonne Universit\'e, Universit\'e Paris Cité, 75005 Paris, France}

\author{Pierre Le Doussal~\orcidlink{0000-0001-6052-1922}}
\affiliation{Laboratoire de Physique de l'\'Ecole Normale Sup\'erieure, CNRS, ENS $\&$ PSL University, Sorbonne Universit\'e, Universit\'e Paris Cité, 75005 Paris, France}

\author{Andrea De Luca~\orcidlink{0000-0003-0272-5083}}
\affiliation{Laboratoire de Physique de l'\'Ecole Normale Sup\'erieure, CNRS, ENS $\&$ PSL University, Sorbonne Universit\'e, Universit\'e Paris Cité, 75005 Paris, France}

\begin{abstract}
We study the Bayesian monitoring of a single particle diffusing along a line, while an observer tracks it continuously using noisy measurements at each spatial point. The average Shannon entropy $\overline{S(t)}$ associated with the
posterior distribution of the particle’s position quantifies the uncertainty that remains after conditioning on the sequence of measurements. We map the problem to the calculation of the moments of the partition functions of directed polymers in $1 + 1$ dimensions. For a constant monitoring rate, the entropy remains bounded for any finite measurement intensity and can be quantified using exact results from the Kardar-Parisi-Zhang equation, providing both the saturation value and the asymptotic behavior for long times. For a monitoring intensity that decreases according to a power law $\sim t^{-\alpha}$, two distinct asymptotic regimes emerge: $\overline{S(t)} \simeq \tfrac12 \ln t$ for
$\alpha > 1/2$ and $\overline{S(t)} \simeq \alpha \ln t$
for $0 < \alpha < 1/2$. Both results are obtained using the replica method, which maps the problem onto an attractive Lieb-Liniger Hamiltonian with a time-dependent coupling: the two regimes can be understood from a perturbative expansion around free diffusion and around the attractive Lieb-Liniger ground state, respectively. We discuss the limiting case $\alpha = 1/2$ and compare the predictions with simulations of discrete Gaussian and log-gamma polymer models.

\end{abstract}

\maketitle

Monitored many-body quantum systems have revealed a new class of nonequilibrium dynamics, in which unitary scrambling~\cite{nahum2017quantum,nahum2018operator,von_Keyserlingk_2018} competes with information extraction by measurements~\cite{sierant2022measurement,ippoliti2021entanglement,potter2022entanglementdynamicsin,lunt2022quantumsimulationusing,sierant2023}. By increasing the measurement rate, a measurement-induced phase transition can emerge, separating a phase in which the entanglement entropy scales with the system volume from a phase where the entropy remains bounded~\cite{li2018quantum,li2019measurementdrivenentanglement,skinner2019measurementinducedphase,chan2019unitaryprojective,szyniszewski2019entanglementtransitionfrom,turkeshi2020measurementinducedcriticality,turkeshi2022measurementinducedcriticality,sierant2022universalbehaviorbeyond, block2022measurement, PhysRevResearch.4.L022066, PhysRevB.107.014306, SciPostPhysCore.5.2.023, PRXQuantum.6.010351}. 
This transition has been investigated intensively, and experimental signatures have been probed on diverse platforms~\cite{google2021,Noel_2022,koh2022experimentalrealizationof,feng2025postselectionfreeexperimental,Sierant2022dissipativefloquet}. 
From a theoretical perspective, monitored systems can be interpreted as disordered systems, with the disorder being given by the intrinsically stochastic record of measurement outcomes, and a major obstacle is that measurement outcomes are not an external disorder (as in standard disordered systems in statistical physics) since their probability is itself state-dependent.

The classical setting is physically different but technically parallel. A stochastic system evolves undisturbed, while noisy measurements performed by an external observer provide partial information that can be used to refine their estimate of the system state by Bayes' theorem~\cite{PhysRevB.106.024305,pizzi2022bridgingthegap}.
Clearly, because the system is classical, the \emph{true} underlying evolution is not affected by the measurement process. However, averaging over measurement records is not a passive disorder average: the probability of a record depends on the inferred state. 
For a diffusive particle, Bayesian monitoring has already been connected, in one spatial dimension, to Kardar-Parisi-Zhang (KPZ) physics at short times~\cite{PhysRevLett.129.260603} and an explicit mapping onto the directed-polymer model provided exact results on the Cayley tree~\cite{kim2024planteddirectedpolymerinferring,gerbino2025measurement}. More recently, the same problem was studied in two and three spatial dimensions~\cite{mcculloch2026bayesian}.
Bayesian measurement-induced criticality has also been investigated in many classical lattice models, including the critical Ising model, by means of replica field-theoretical approaches~\cite{nahum2025bayesian,wiese2026bayesian}.
\begin{figure}
\centering
\includegraphics[width=\linewidth]{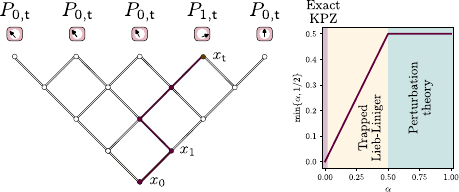}
    \caption{\textit{Left:} Model of Bayesian monitoring in one spatial dimension. \textit{Right:} Rate of production of Shannon entropy as a function of the exponent $\alpha$ controlling the asymptotic decay of the monitoring rate.}
    \label{fig:model}
\end{figure}

In this Letter we consider a single particle diffusing on the line. The observer cannot access the physical trajectory, but receives noisy local observations and updates a posterior distribution $\{p_j^{(\t)}\}_{j\in \mathbb Z}$. We use the Shannon entropy
\be
S_\t : = -\sum_{j\in \mathbb Z} p_j^{(\t)} \ln p_j^{(\t)}
\label{eq:entropydef}
\ee
as the central observable, and we consider its average $\average{S_\t}$ over the realizations of the measurement outcomes: an unbounded growth of $\average{S_\t}$ corresponds to a broadening of the posterior distribution, whereas constant entropy signals (approximate) detection of the particle's position.
Without measurements one recovers diffusion, hence a logarithmic growth  $S_\t=\frac12\ln \t+O(1)$, which provides the maximal asymptotic behavior compatible with the underlying diffusive dynamics. For constant nonzero values of the monitoring strength, we show that the average entropy is always $O(1)$ at late times. However, as the particle continues to spread, monitoring its position throughout the entire spatial domain becomes increasingly demanding; therefore, we also consider the relevant case in which the monitoring rate decreases over time. For decreasing monitoring strength, asymptotically parameterized by a power-law $\sim \t^{-\alpha}$, the average entropy exhibits two regimes,
\be
\average{S_\t} =
\begin{cases}
\alpha \ln \t + O(1), & 0\leq\alpha\leq1/2,\\
\frac12 \ln \t + O(1), & \alpha>1/2,
\end{cases}
\label{eq:mainresult}
\ee
with a marginal case at $\alpha=1/2$, see Fig.~\ref{fig:model}.
This result thus shows that the asymptotic scaling of the entropy depends on the balance between diffusive spreading and the monitoring rate. As we will see, in addition to the leading term, the $O(1)$ corrections also exhibit universal contributions in this process.

\textit{Discrete model and the directed polymer.}---We start from the lattice formulation introduced in Ref.~\cite{gerbino2025measurement}. We consider a particle undergoing a random walk on the one-dimensional lattice, with each trajectory identifying a path in the 1+1-dimensional lattice. Let $p_j^{(\t)}$ denote the probability of occupation at site $-\t\leq j \leq \t$ after $\t$ steps, estimated by the observer. 
The particle starts from the origin, therefore $p_j^{(\t=0)}=\delta_{j,0}$. 
At each time-step, the particle hops left or right with equal probabilities and $p_j^{(\t)}$ propagates to the neighboring sites $j\pm 1$. Then, the observer performs a noisy measurement of the particle position and updates the probability distribution $\mathbf p^{(\t+1)} := \{p_j^{(\t+1)}\}_j$. Essentially, at each site $j$  and time $\t$, the observer receives a random variable $a_j^{(\t)}$ whose distribution is $P_{1,\t}$ if the particle is present at that site and $P_{0,\t}$ otherwise. Notice that we consider the distributions $P_{s,\t}$ to be explicit functions of time. The model is depicted in Fig.~\ref{fig:model}. 
Therefore, the probability of the full set of observations $\mathbf a^{(\t)}:=\{a_j^{(\t)}\}_j$ at a given time $\t$ conditioned on the particle being at site $j$ -- the likelihood -- is given by 
\begin{equation}
\label{eq:pafromj}
    P(\mathbf{a}^{(\t)}|j) = \frac{P_{1,\t}(a_j^{(\t)})}{P_{0,\t}(a_j^{(\t)})} \prod_{j'\in \mathbb Z}P_{0,\t}(a_{j'}^{(\t)}) \,.
\end{equation}
Because $\mathbf p^{(\t)}$ depends on the measurement outcomes, we call it the posterior distribution. 
Combining the diffusive step  
with Bayes' theorem, we get the update rule 
\be
p_j^{(\t+1)} =
\frac{P(\mathbf a^{(\t+1)}|j)\,[p_{j-1}^{(\t)}+p_{j+1}^{(\t)}]}{\sum_{j'\in \mathbb Z} P(\mathbf a^{(\t+1)}|j')\,[p_{j'-1}^{(\t)}+p_{j'+1}^{(\t)}]}\,.
\label{eq:bayesdiscrete}
\ee
In order to eliminate the denominator above, we introduce the unnormalized amplitudes $z_j^{(\t)}$, satisfying the discrete stochastic heat equation (SHE) with initial condition $z_j^{(\t=0)}=\delta_{j,0}$:
\be 
\small 
z_j^{(\t+1)}=B_{\t+1}(a_j^{(\t+1)})\left[z_{j-1}^{(\t)}+z_{j+1}^{(\t)}\right],
\,\,
B_\t(a)=\frac{P_{1,\t}(a)}{2P_{0,\t}(a)} .
\label{eq:discreteSHE}
\ee
The occupation probability is retrieved by normalization $p_j^{(\t)} = z_j^{(\t)}/Z^{(\t)}$, with $Z^{(\t)}=\sum_{j \in \mathbb Z}z_j^{(\t)}$.
Iterating Eq.~\eqref{eq:discreteSHE} allows expressing $z_j^{(\t)}$ as a sum over paths from $0$ to $j$, each weighted by the product of the likelihood ratios $B_{\t'}(a)$ encountered along the path.
Interpreting $B_{\t'}(a)$ as a Boltzmann weight, $z_j^{(\t)}$ is exactly the point-to-point partition function of a directed polymer (DP) on a square lattice with a random onsite potential, while $Z^{(\t)}$ is the point-to-line partition function.
We treat two examples: Gaussian measurements $P_{s,\t} = \mathcal N(s\beta_\t,1)$, 
and $P_{s,\t}(a)\propto e^{-1/ a} a^{-1-(\gamma_\t-s)}$, i.e.,  the outcomes are inverse-Gamma-distributed. 
These examples correspond to the Gaussian and the log-gamma polymer respectively~\cite{calabrese2010free,barraquand2020stochastic,Le_Doussal_2016}. 
However, according to Eq.~\eqref{eq:pafromj}, the distribution of the measurement outcomes -- the onsite potential in the DP framework -- depends on the real path followed by the monitored particle. 
By averaging uniformly over such paths,  one arrives at the following identity
\begin{equation} \label{eq:Born}
    \average{F[\mathbf{p}^{(\t)}]} = 
    \average{F\left[ \frac{\mathbf{z}^{(\t)}}{Z^{(\t)}}\right]Z^{(\t)}}_0 \,,
\end{equation}
where $F$ is any functional of the inferred probabilities. Here and throughout the text, $\average{\cdots}$ denotes the average over the outcomes $\{\mathbf a^{(\t)}\}_\t$: by Eq.~\eqref{eq:pafromj}, their measure is induced by the uniform distribution of paths of the monitored particle; in contrast, the subscript $\average{\dots}_0$ indicates the simpler process where all the outcomes are i.i.d. distributed with $P_{0,\t}$~\cite{gerbino2025measurement}. 
Consistently with the above identity, we have $\average{B_\t(a)}_0=1/2$ and $\average{Z^{(\t)}}_0=1$. Using Eq.~\eqref{eq:entropydef} and \eqref{eq:Born}, we have 
\begin{equation} \label{eq:Sdiscrete}
    \average{S_\t} = \average{Z^{(\t)}\ln Z^{(\t)}}_0 - \sum_{j \in \mathbb Z} \average{z_j^{(\t)} \ln z_j^{(\t)}}_0 \,.
\end{equation}
Unlike in the case of the Cayley tree~\cite{gerbino2025measurement}, both terms on the r.h.s. of the identity for $\average{S_\t}$ are non-trivial, as in 1+1 dimensions the partition function $z_j^{(\t)}$ at a specific site $j$ does not identify a unique path. 

\textit{Continuum limit and replica formulation.}---Under diffusive scaling~\cite{brunet2000influence,Le_Doussal_2016,barraquand2020stochastic} $x = j\Delta x$, $t = \t \Delta t$ with $\Delta t = \Delta x^2/2\to 0$,
and a weak-noise limit of monitoring ($P_{0, \t}\simeq P_{1,\t}$, see Sec.~A in End Matter (EM)), Eq.~\eqref{eq:discreteSHE} converges to the SHE
\be
\partial_t z(x,t)=\partial_x^2 z(x,t)+\sqrt{2c(t)}\,\xi(x,t)z(x,t),
\label{eq:SHE}
\ee
in the continuous variables $x,t$,
with the standard white noise $\overbar{\xi(x,t)\xi(x',t')}=\delta(x-x')\delta(t-t')$, and initial condition $z(x,t=0)=\delta(x)$.
For clarity, we use the notation $\overline{(\dots)}$ in the continuum, equivalent to the unbiased average $\average{\cdots}_0$ introduced above on the lattice. The effective noise variance $c(t)$ is the continuum measure of the monitoring strength: it is fixed by the information carried by a single round of measurements, i.e., by the Kullback-Leibler divergence between the two outcome distributions, $c(t):= \lim_{\Delta t \to 0} 2\sqrt{2/\Delta t}\ D_{\rm KL}(P_{1,\t} || P_{0,\t})$, which we assume to be finite and time dependent in general; it represents a tunable parameter controlling the monitoring strength in the continuum limit (see Eq.~\eqref{eq:cdef} in EM).
We highlight that the variable $h(x,t):=\ln z(x,t)$ obeys a KPZ equation with time-dependent noise amplitude~\cite{barraquand2020stochastic,halpin1995kinetic,comets2017directed,kamenev2023field,fradkin2013field}. The posterior is
\be
p(x,t)=\frac{z(x,t)}{Z(t)},\quad Z(t)=\int_{\mathbb R} d x\,z(x,t), \quad 
\overline{Z(t)}=1 ,
\label{eq:pfromz}
\ee
and the average Shannon entropy \eqref{eq:Sdiscrete} becomes
\be 
\overbar{S(t)}=\overbar{Z(t)\ln Z(t)}-\int_{-\infty}^\infty d x\,\overbar{z(x,t)\ln z(x,t)},
\label{eq:replicaentropy}
\ee
up to a lattice-dependent additive constant $-\ln 2\Delta x$, which is subtracted throughout (see Eq.~\eqref{eq:Sdiv} in the EM).

The terms in Eq.~\eqref{eq:replicaentropy} admit the replica representations
\begin{align}
S_{\rm F}(t) := \partial_{n=1} \overbar{Z(t)^n}, \;
S_{\rm D}(t) := \partial_{n=1} \int_{\mathbb R} dx \  \overbar{z(x,t)^n}.
\label{eq:replicamoments}
\end{align}
In the KPZ language, they correspond to moments of the partition function with ``flat'' and ``droplet'' initial conditions respectively. 
For integer $n$, the noise average of the moments $\overbar{\prod_{i=1}^n z(x_i,t)} =:\langle \xvec \ket{\Psi(t)}$ can be expressed via the imaginary-time Schr{\"o}dinger formalism introducing 
\begin{equation}
\partial_t \ket{\Psi(t)} = -\hat{H}_{\rm LL}(c(t))
\ket{\Psi(t)}\,,  \label{eq:schrodinger}
\end{equation}
with the attractive Lieb-Liniger (LL) Hamiltonian 
\begin{equation}
    \hat{H}_{\rm LL}(\bar c) := -\sum_{i=1}^n\nabla^2_i - 2\bar c \sum_{1\leq i<j\leq n} \delta(x_j-x_i) \,.
    \label{eq:LL}
\end{equation}
In our case $\bar c = c(t)$ and the initial condition is $\langle \xvec \ket{\Psi(t=0)}=\prod_{i=1}^n\delta(x_i)$.  
The disorder average therefore couples the replicas attractively, with a time-dependent coupling constant $c(t)$ fixed by the noise variance of Eq.~\eqref{eq:SHE}. 
Importantly, in the replica limit $n \to 1$, the extensive contributions to the flat and droplet terms cancel in the difference \eqref{eq:replicaentropy}, consistently with the fact that the Shannon entropy is always subextensive in $t$ (even in the absence of monitoring).
From a statistical physics perspective, the entropy problem differs from the usual polymer free-energy problem because the derivative is taken at $n=1$ rather than at $n=0$: one must keep track of the normalization factor $Z(t)$ arising from the ``Born'' rule Eq.~\eqref{eq:Born}~\cite{zhang2021emergentreplica,li2021statisticalmechanicsof,giachetti2023elusivephasetransitionreplica,deluca2025universality}. 

\begin{figure*}[t]
\centering
\includegraphics[width=0.66\columnwidth]{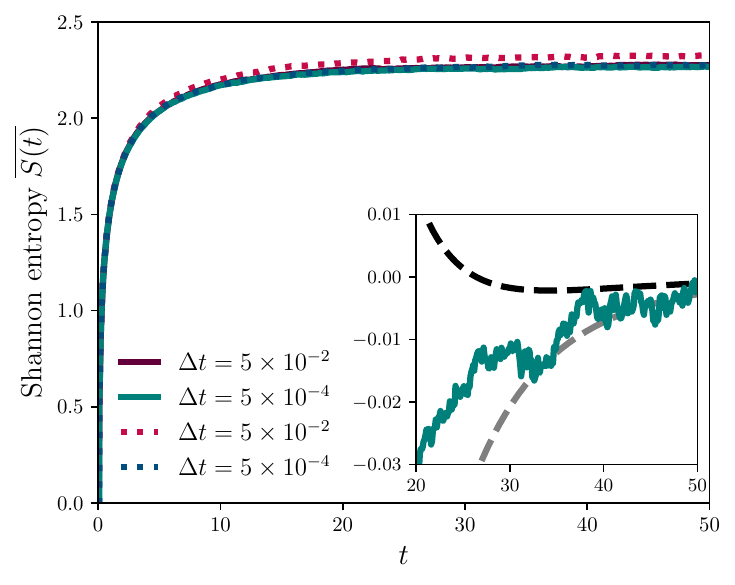}
\hfill
\includegraphics[width=0.66\columnwidth]{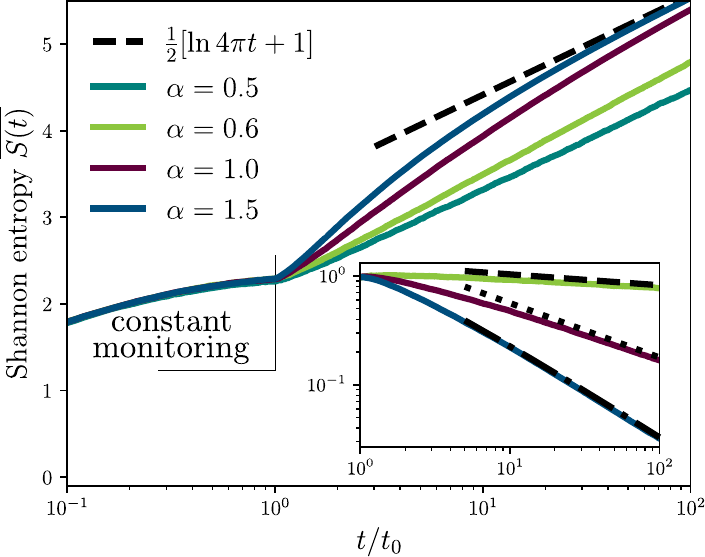} 
\hfill
\includegraphics[width=0.66\columnwidth]{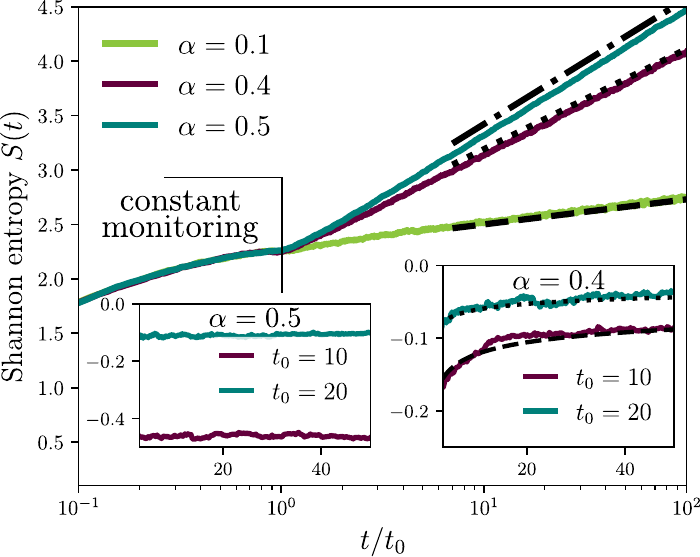} 
\caption{\textit{Left:} Saturation to a constant entropy for $\alpha=0$, $\bar c=1$. Different curves show different values of the time-step $\Delta t$ in the diffusive scaling, and for two choices of the measurement distributions $P_{0,1,\t}$: inverse-Gamma for solid lines, Gaussian for dotted lines. Inset: approach to saturation, with the constant $S_\infty$ Eq.~\eqref{eq:alphazero} removed. 
The dashed black line is the theoretical Eq.~\eqref{eq:approach}, dashed gray line is \eqref{eq:approach} to $O(t^{-3/2})$.
\textit{Center:} Diffusive entropy growth of numerical data for various $\alpha\geq 1/2$ ($t_0=20$, $c_0=1$, $\Delta t=5\times 10^{-2}$), compared to unmonitored entropy (dashed line). Inset: numerical $\s[t] - \frac12(\ln4\pi t+1)$ compared to the perturbative corrections Eqs.~(\ref{eq:entroalphalarge}, \ref{eq:entro32}). \textit{Right:} $\alpha$-dependent entropy growth of numerical data for various $0< \alpha\leq1/2$ ($t_0=20$, $c_0=1$, $\Delta t=5\times 10^{-3}$), compared to $-\ln c(t) +S_\infty$ (black lines). Inset: numerical $\s[t]+\ln c(t)$, compared to the $O(t^{2\alpha-1})$ correction in Eq.~\eqref{eq:alphaLT} (black lines), for $\alpha=0.4$.
See~\cite{supplnote} for a discussion of the numerical analysis. }
\label{fig:numerics}
\end{figure*}

\textit{Constant monitoring.}---For constant $c(t)=\bar c>0$ Eq.~\eqref{eq:schrodinger} reduces to the evolution with the time-independent attractive LL model, and exact KPZ results become available through the DP correspondence. One finds that typically the posterior $p(x,t)$ is \emph{localized}, i.e.,  it has an $O(1)$ Shannon entropy at large times. One can always reabsorb the constant $\bar c$ in Eq.~\eqref{eq:SHE} by a rescaling of space and time. Accordingly, since $z(x,t)$ has to be normalized (cf. Eq.~\eqref{eq:pfromz}), this implies that $S_{\rm D}(t)$ takes a constant shift $\ln \bar c$; consequently
\be
\overbar{S(t)}= \underbrace{1+\gamma_{\rm E}+\ln 2}_{S_\infty} - \ln \bar c+ \delta \mathcal S(\bar c^2 t) \,, 
\label{eq:alphazero}
\ee
with the function of the rescaled time $\bar c^2 t$:
\begin{equation} \label{eq:approach}
\small
   \delta  \mathcal S(u) = - 4 e^{-\frac{u}{16}} \frac{\pi^{3/2}}{u^{3/2}} \left[ 1 +\frac{a_1}{u} + \frac{a_2}{u^{3/2}} + O(u^{-2}) \right] + \ldots ,
\end{equation}
with $a_1=-2(12 + \pi^2)$ and $a_2 = 16\pi^{3/2}$ and $\ldots$ contains subleading exponential corrections.
In the following, we set $\bar c=1$ for simplicity; Eq.~\eqref{eq:alphazero} can be used for the general case.
In the derivation of Eq.~\eqref{eq:alphazero}, we do not exploit the replica representation \eqref{eq:replicamoments}, but consider instead the full distributions of the polymer partition functions $z$ and $Z$. In particular, the computation of the Shannon entropy is conveniently expressed from appropriate generating functions of the moments of $z(0,t)$ and $Z(t)$, that are known exactly. It is convenient to introduce the rescaled point-to-point partition sum $\hat z(t):=\sqrt{4\pi t}\;z(0,t)$, normalized so that $\overline{\hat z(t)}=1$, with $\hat z(0)=1$; it plays for the droplet geometry the same role that $Z(t)$, with $\overline{Z(t)}=1$, plays for the flat one. 
Notice that 
it is sufficient to have access to
the distribution at the origin $x=0$ as, by statistical tilt symmetry~\cite{supplnote},
one can remove the $x$-dependence and arrive at
\begin{equation}\label{eq:SDhat}
S_{\rm D}(t) =  \overline{\hat z(t) \ln \hat z(t)} -\frac{1}{2} [1 + \ln(4\pi t)] \,. 
\end{equation}
For the droplet geometry, one has the exact generating function for $\hat z$, expressed in terms of the Fredholm determinant~\cite{sasamoto2010one,calabrese2010free,Le_Doussal_2016}
\begin{align} \label{eq:gendroplet}
    g_{\rm D}(s,t) :&= \overline{\exp\left[-e^{-(s-\theta(t))} \ \hat z(t) \right]} \\
    &= \det\left( \mathbb{I} -  P_{st^{-1/3}}  K_t P_{st^{-1/3}}\right) \,, \nonumber
\end{align}
with the shift $\theta(t) := \frac{t}{12}- \frac12\ln 4\pi t$, and the Airy kernel 
\begin{equation} \label{eq:airy}
K_{t}(r, r') = \int_{-\infty}^\infty dv \frac{\text{Ai}(r+v)\text{Ai}(r'+v)}{1 + e^{- t^{1/3}v}} \,,
\end{equation}
where $\text{Ai}$ indicates the Airy function, and $P_b$ is the projector onto the interval $[b,+\infty[$.
For the flat geometry, the generating function \begin{equation} \label{eq:genflat}
    g_{\rm F}(s,t) := \overline{\exp\left( -e^{-(s-t/12)} Z(t)\right) }
\end{equation}
admits a more involved expansion in terms of a Fredholm Pfaffian~\cite{le2012kpz}. 
By means of the generating functions $g_{\rm D/F}$, we can express the Shannon entropy terms as~\cite{gerbino2025measurement} 
\begin{equation} \label{eq:fromZtoG}
\begin{split}
    & \overline{\hat{z}(t) \ln \hat{z}(t)} =\int_{\mathbb R} ds\ e^s\ \left[g_{\rm D}(s+\theta(t),t) -e^{-e^{-s}}\right] \,,\\
    &S_{\rm F}(t) = \int_{\mathbb R} ds \ e^{s}[ g_{\rm F}(s+t/12,t) - g_{\rm F}(s,0)]\,.
\end{split}
\end{equation}
These expressions are exact and encode the full time-dependence of the Shannon entropy, but they are involved and rather implicit. Since we only aim at the long-time asymptotics, the complete expressions of the determinants are not needed for the evaluation of the integrals~\eqref{eq:fromZtoG}: it is enough to compute the first two terms of their large-time expansion~\cite{supplnote}, both for the flat and the droplet geometry. At the leading
order,
$S_{\rm F/D}(t) = t/6 + c_{\rm F/D} + o(1)$: as anticipated, the $\propto t$ contributions cancel in Eq.~\eqref{eq:replicaentropy}, while the constants lead to $\s[t] \to c_{\rm F} - c_{\rm D} = S_\infty$. Higher orders in the expansion of Eqs.~(\ref{eq:gendroplet}, \ref{eq:genflat}) give the approach to the stationary value Eq.~\eqref{eq:alphazero}~\cite{supplnote}, checked numerically in Fig.~\ref{fig:numerics} (left).

From the inference point of view, Eq.~\eqref{eq:alphazero} means that any fixed nonzero monitoring rate is sufficient to localize the posterior, and there is no threshold monitoring strength below which a diverging entropy is recovered.
This should be contrasted both with the one-dimensional quantum many-body case~\cite{PhysRevX.9.031009} and with the monitored random walk on the Cayley tree~\cite{kim2024planteddirectedpolymerinferring,gerbino2025measurement}, where a genuine transition can occur at constant rate. 

\textit{Time-dependent power-law monitoring.}---We now let the effective monitoring strength $c(t)$, i.e., the noise variance in Eq.~\eqref{eq:SHE}, be a non-constant function of time. We set $c(t=0)=1$ for simplicity, and, while we do not assume it to be a specific function at early times, we specialize to the case
\be
c(t)\simeq c_0 \left( \frac{t}{t_0} \right)^{-\alpha} \,,\quad t\geq t_0 \,,
\label{eq:ct}
\ee
with $c_0$ a small parameter and $t_0$ arbitrary.

The structure of the problem is controlled by a single dimensionless coupling. In Eq.~\eqref{eq:LL} the replica attraction $\bar c  = c(t)$ binds the replicas over a length $\sim 1/c(t)$, to be compared with the diffusive scale $\sqrt t$ explored at time $t$; the relevant coupling is therefore
\be \label{eq:gdimless}
g(t) := c(t)\sqrt{t} \simeq c_0\sqrt{t_0}\ (t/t_0)^{\frac12 -\alpha}\,.
\ee
For $\alpha>1/2$ one has $g(t)\to0$: the replicas are asymptotically free, the interaction can be treated as a perturbation of free diffusion, and monitoring is unable to affect the leading diffusive growth of the entropy. For $\alpha<1/2$ one has $g(t)\to\infty$: the replicas are tightly bound and the natural starting point of the expansion is the LL bound state itself. 
The marginal case $\alpha=1/2$ is the one where $g(t)=g_*:=c_0\sqrt{t_0}$ stays constant, so that no expansion is available and the result may depend on $g_*$.

\textit{Case $\alpha > 1/2$.}---Starting from the state $\ket{\Psi(t_0)}$ generated by the evolution~\eqref{eq:schrodinger} at times $t<t_0$, we organize the large-time expansion using powers of $c_0$ in the interaction picture with respect to the free Hamiltonian $\hat H_0 := -\sum_{i=1}^n \nabla^2_i$. Up to first order, we have $\s[t] = S^{(0)}(t) + c_0 S^{(1)}(t) + O(c_0^2)$ (see Sec.~B in EM).
The zeroth-order describes $n$ free diffusing replicas and gives the unmonitored entropy $S^{(0)}(t)= \tfrac12[1+\ln4\pi t] + O(t^{-1})$. The
$O(t^{-1})$ correction is
a consequence of the short-time dynamics $t \in [0, t_0]$ and in this formalism is determined by the precise shape of $\ket{\Psi(t_0)}$ which we assume unknown. 
Instead, $S^{(1)}(t)$ is controlled by the time-integrated probability that two replicas meet at $t' \in (t_0,t)$, weighted by $c(t')$: it differs between the flat and droplet geometries because in the latter the replicas are conditioned to meet again at time $t$. At large time $t$, $S^{(1)}(t)$ can be expressed explicitly neglecting the short-time broadening, i.e.,  $\langle \xvec | \Psi(t_0)\rangle \approx 
\langle \xvec | \Psi(0) \rangle= \prod_{i=1}^n \delta(x_i)$.
After differentiation in $n$, one has
\begin{equation} \label{eq:firstterm}
    S^{(1)}(t) \stackrel{t \gg t_0}{\sim} -\frac{t_0^\alpha}{\sqrt{8\pi}}\int_{t_0}^{t} \frac{dt'\ t'^{-\alpha}}{\sqrt{t'-t_0}}\left[\sqrt{\frac{t-t_0}{t-t'}}-1\right] \,.
\end{equation}
The term in the square bracket follows from the difference between the two geometries (see Eq.~\eqref{eq:CFCD} in EM). 
For $1/2< \alpha < 3/2$, expanding the integral \eqref{eq:firstterm} at $t\to \infty$,
the Shannon entropy at first order in $c_0$ reads
\begin{equation} \label{eq:entroalphalarge}
    \s[t] = \frac12[1+\ln4\pi t]
    - \frac{c_0 t_0^\alpha}{\sqrt{8\pi}}\ I(\alpha) \ t^{\frac12-\alpha} + O(t^{-1})\,, 
\end{equation}
with $I(\alpha)$ a positive factor (Eq.~\eqref{eq:Ialpha_min} in  EM), and $\s[t]$ 
therefore approaches $S^{(0)}$ from below, with a relative correction $O(t^{1/2 - \alpha}) = O(g(t))$ that vanishes faster the larger $\alpha$: the perturbative expansion around free diffusion is self-consistent for every $\alpha>1/2$.
One has $I(\alpha \to 3/2) \to \infty$ and the integral~\eqref{eq:firstterm} features $\log$-corrections, leading to
\begin{equation} \label{eq:entro32}
    \s[t] = \frac12[1+\ln4\pi t] - \frac{c_0 t_0^{3/2}}{4\sqrt{2\pi}\, t} \ln \frac{t}{t_0}+O(t^{-1}) \,.
\end{equation}
Finally, for $\alpha>3/2$, $S^{(1)}(t)\sim t^{-1}$ asymptotically, and competes with the corrections generated by the unknown initial state $\ket{\Psi(t_0)}$.
Therefore, up to the single non-universal amplitude $c_0t_0^\alpha$, the expressions (\ref{eq:entroalphalarge}, \ref{eq:entro32}) are universal: they do not depend on the monitoring profile before $t_0$, see Fig.~\ref{fig:numerics} (center) for a numerical check. 

\textit{Case $0<\alpha<1/2$.}---In this regime, the previous expansion breaks down, and the corrections \eqref{eq:firstterm} grow as $g(t) \propto t^{1/2-\alpha} \to\infty$.
Following Ref.~\cite{barraquand2020stochastic}, we trade the time dependence of the coupling for a time-dependent harmonic potential. For a generic $c(t)$, the coordinate $y=c(t)x$, the time $\tau(t)=\int_0^t d s\, c(s)^2$ and a rescaling of $z(x,t)$ into a new field $\zz(y,\tau)$ (Eq.~\eqref{eq:mapping} in the EM) turn Eq.~\eqref{eq:SHE} into a SHE with \emph{time-independent} noise strength in a harmonic trap,
\be
\partial_\tau \zz=\partial_y^2\zz-\frac{A(\tau)}{2}\,y^2\zz+\sqrt{2}\,\xi(y,\tau)\zz \;,
\label{eq:SHEharmonic}
\ee
of amplitude $A(\tau)=c''(\tau)/(2c(\tau))$, where primes denote $d/d\tau$ and, with a slight abuse of notation, $c(t)$ and $c(\tau)$ denote the same function of different arguments.
The evolution of the modified moments $\overbar{\prod_{i=1}^n \zz(y_i,\tau)} =:\langle \yvec \ket{\check \Psi(\tau)} $ is governed by a Schr{\"o}dinger equation analogous to~\eqref{eq:schrodinger}:
\begin{equation}
    \partial_\tau \ket{\check \Psi(\tau)} = - \left[ \hat H_{\rm LL}(1) + \frac{A(\tau)}{2}\sum_{i=1}^n y_i^2 \right] \ket{\check \Psi(\tau)}\,,
\label{eq:LLharmonic}
\end{equation}
where $\hat{H}_{\rm LL}(1)$ is the attractive LL Hamiltonian~\eqref{eq:LL} at fixed coupling $\bar c=1$, now written in the $\yvec$ coordinates. The initial condition is $\langle \yvec \ket{\check \Psi(\tau=0)} = \prod_{i=1}^n \delta(y_i)$. 

Since the trap is quadratic, the center of mass $R=\tfrac1n\sum_{a=1}^n y_a$ decouples from the gap coordinates $\tilde y_i=y_{i+1}-y_i$, $1\leq i \leq n-1$, and the propagator factorizes,
\be 
\langle \yvec \ket{\check \Psi(\tau)} 
=K_n(R;\tau)\,
k_n(\tilde y_1,\dots,\tilde y_{n-1};\tau) \,.
\label{eq:COMREL}
\ee
The center-of-mass piece is a Gaussian one-body propagator, solvable for any $A(\tau)$~\cite{supplnote}, but it drops out of the entropy altogether: writing $\sum_{a=1}^n y_a^2=nR^2+\f(\tyvec)$, with $\f$ quadratic in the relative coordinates only, it enters both moments \eqref{eq:replicamoments} through an exact identity (Eq.~\eqref{eq:COM} of the EM) which generalizes the conservation of the norm $\overline{Z(t)}=1$ to $n>1$. Integrating over $R$, we are thus left with the relative propagator $k_n$ only (see Eq.~\eqref{eq:replicamapped} in the EM),
\begin{subequations}
\label{eq:momentsmapped}
\begin{align}
& \overline{Z^n(t)} = c(\tau)^{\frac{1-n}{2}} \int_{\mathbb R^{n-1}} d^{n-1} \tilde y \ e^{\frac{c'(\tau)}{4c(\tau)} \f(\tilde{\mathbf y})} k_n(\tilde{\mathbf y} ,\tau) \,, 
\label{eq:ZnTmain}
\\
& \int_{\mathbb R} d x \ \overline{z^n(x,t)} = c(\tau)^{\frac{n-1}{2}} \ k_n(0,\tau) \,.
\label{eq:zntmain}
\end{align}
\end{subequations}
Plugging these expressions in Eqs.~(\ref{eq:replicaentropy}, \ref{eq:replicamoments}), and writing $k_n^{\rm F}(\tau) := c(\tau)^{\frac{n-1}{2}} \overline{Z^n(t)}$, we obtain
\begin{equation}
\label{eq:Sdiffder}
   \s[t] = -\ln c(\tau) + \partial_n \left.(k_n^{\rm F}(\tau) - k_n(0,\tau))\right|_{n=1}\,.
\end{equation}
These equations are valid for arbitrary $c(t)$; for $c(t)$ of the shape \eqref{eq:ct}, one has $\tau(t) = \tau_0 + \frac{c_0^2 t_0}{1-2\alpha}\left(\frac{t}{t_0}\right)^{1-2\alpha}$, with $\tau_0$ depending on the unknown early-time shape of $c(t)$ and
\begin{equation} \label{eq:potentialdef}
    A(\tau) := \frac \alpha 2\frac{1-\alpha}{(1-2\alpha)^2}\frac{1}{(\tau-\tau_0)^2} \,.
\end{equation}
Note that in the regime $t > t_0$, one has $\tau > \tau_0$, so the apparent divergence $\tau \to \tau_0$ is never relevant.  
Here, we focus on the regime $0<\alpha<1/2$, when 
$\tau(t)\stackrel{t\to\infty}{\to}\infty$, $A(\tau)\stackrel{t\to\infty}{\to}0$, and the trap becomes negligible at large times. Thus, the evolution \eqref{eq:LLharmonic} asymptotically reduces to the pure LL one, which projects onto the ground state $\ket{\Psi^{\rm GS}_n}$ of $\hat H_{\rm LL}(1)$; a perturbative treatment of the weakening trap~\cite{supplnote} shows that 
\be \label{eq:GSproj}
k_n(\tyvec,\tau) = \kappa_n(\tau)\, e^{-\tau E_n^{\rm GS}}\big(\psi_n(\tyvec) + O(\tau^{-2})\big)\,,
\ee
where $\psi_n (\tyvec) := \bra{\tyvec}\Psi^{\rm GS}_n\rangle\braket{\Psi^{\rm GS}_n|\check\Psi(\tau=0)}$ (explicitly given in Eq.~\eqref{eq:psin} of the EM), and the amplitude $\kappa_n(\tau)$ depends on the short-time evolution. In~\cite{supplnote}, we derive its large-$\tau$ asymptotic expansion, but here we only need that $\lim_{n\to1}\kappa_n(\tau)=1$ by normalization.
Indeed, the factor $\kappa_n(\tau)e^{-\tau E^{\rm GS}_n}$ multiplies the flat and the droplet term alike:
in the difference \eqref{eq:Sdiffder} only its trivial value at $n=1$ is relevant (see Eq.~\eqref{eq:entroexplicit}). The only subleading term that distinguishes the two geometries is the Gaussian factor in Eq.~\eqref{eq:ZnTmain}, whose exponent also vanishes asymptotically:
$\exp[c'(\tau) \f(\tyvec)/(4c(\tau))] =1-\tfrac{\alpha}{4}(1-2\alpha)^{-1}(\tau-\tau_0)^{-1} \f (\tyvec) + O(\tau^{-2})$. 
The integrals over $\tilde \yvec$ can be computed from the explicit form of the ground state (see Eq.~\eqref{eq:psimoments} in EM) which provides an analytic continuation in $n$ leading to
\be 
\s[t]= - \ln c(t) +S_\infty - \frac{4 \alpha \zeta(3)}{c_0^2 t_0^{2\alpha}}
\frac{1}{t^{1-2\alpha}} +
O(t^{2(2\alpha-1)}) \,,
\label{eq:alphaLT}
\ee
where one can make explicit $-\ln c(t) = \alpha \ln \tfrac{t}{t_0}-\ln c_0$.
The comparison with numerical data for the entropy is in Fig.~\ref{fig:numerics} (right). Note that the constant $S_\infty$ of Eq.~\eqref{eq:alphazero} obtained at $\alpha=0$ is reproduced.  

A similar mechanism gives the leading order of the entropy in the marginal case $\alpha=1/2$. Here $\tau(t)$ still diverges, but only logarithmically, and both the trap amplitude and the exponent of the Gaussian factor in Eq.~\eqref{eq:momentsmapped} are finite, $A=(8g_*^4)^{-1}$ and $c'/(4c)=-(8g_*^2)^{-1}$, controlled by the marginal coupling $g_*=c_0\sqrt{t_0}$ alone. The relative propagator is then projected onto the ground state of the LL model in a \emph{static} harmonic trap. While the $n$-dependence of $k_n(0,\tau)$ and of the corresponding integral is not known in this case, using $k_1(\tilde y,\tau)=1$ and $-\ln c(t)=\frac12\ln t-\ln g_*$ we still obtain
$\s[t] = \frac12\ln t + \Phi(g_*)$, with $\Phi$ a nonuniversal function of the marginal coupling. This is the only point where the $O(1)$ term retains a memory of the amplitude of the protocol, and there is no reason for $\Phi(g_*)$ to match the $\alpha\to1/2^\pm$ limits of Eqs.~(\ref{eq:entroalphalarge},~\ref{eq:alphaLT}): those are attained only after the crossover scale set by $g_*$, which diverges as $\alpha\to1/2$. 

\textit{Discussion.}---Bayesian monitoring of a diffusive $1d$ particle furnishes a solvable inference problem in which the replica structure is essential for averaging over the measurement records. The directed-polymer mapping makes it possible to derive nontrivial late-time predictions directly for the observer's entropy. Remarkably, constant monitoring always keeps the uncertainty finite; therefore, the observer is always able to (on average) localize the particle. Conversely, a power-law decay in time of the monitoring strength produces a threshold at $\alpha=1/2$, separating a regime where the entropy growth is reduced by measurements from a diffusion-dominated regime where measurements are ineffective; the two regimes are the two possible fates of the dimensionless coupling \eqref{eq:gdimless}. In both of them the replica construction also shows that the late-time behavior is controlled by only two non-universal parameters, $t_0$ and $c_0$, entering through the single combination $c_0t_0^\alpha$: any earlier evolution is eventually washed out by the diffusive growth of the random-walk entropy.

Several extensions are natural. First, the marginal point $\alpha=1/2$, which cannot be accessed perturbatively, deserves a dedicated analysis in the trapped LL formulation. Moreover, higher dimensions should be especially interesting, as a genuine monitoring transition may already occur at constant rate~\cite{nahum2025bayesian,wiese2026bayesian, mcculloch2026bayesian} and the interplay with a time-dependent monitoring can largely enrich the phase diagram. It would also be worthwhile to investigate optimized protocols aiming at locating the particle with reduced effort taking advantage of previous observations.

\begin{acknowledgements}
\textit{Acknowledgements.---}  G.G. acknowledges the support of the European MSCA Grant 101152898 (DREAMS). PLD acknowledges support of the ANR Grant No. ANR-23-CE30-0020-01
EDIPS.

\end{acknowledgements}

\bibliography{references}

@article{calabrese2010free,
  title={Free-energy distribution of the directed polymer at high temperature},
  author={Calabrese, Pasquale and Le Doussal, Pierre and Rosso, Alberto},
  journal={Europhys. Lett.},
  volume={90},
  number={2},
  pages={20002},
  year={2010},
  url = {https://iopscience.iop.org/article/10.1209/0295-5075/90/20002/pdf?casa_token=nZAu4yQVQVsAAAAA:1BIm0LTGrbHYaiBEWEToIvsYpJn9yRtpuC1WPSyoCkGoFRPJcK6ccnd7R06Px9Amagp4AYpfrvbUk26Pn438Tsdu24L9-w},
  publisher={IOP Publishing}
}

@article{sasamoto2010one,
  title={One-dimensional Kardar-Parisi-Zhang equation: an exact solution and its universality},
  author={Sasamoto, Tomohiro and Spohn, Herbert},
  journal={Phys. Rev. Lett.},
  volume={104},
  number={23},
  pages={230602},
  year={2010},
  url={https://journals.aps.org/prl/abstract/10.1103/PhysRevLett.104.230602},
  publisher={APS}
}

@article{barraquand2020stochastic,
  title={Stochastic growth in time-dependent environments},
  author={Barraquand, Guillaume and Le Doussal, Pierre and Rosso, Alberto},
  journal={Phys. Rev. E},
  volume={101},
  number={4},
  pages={040101},
  year={2020},
  publisher={APS}
}

@misc{supplnote,
 note = {see Supplemental Material for additional details}
}

@article{halpin1995kinetic,
  title={Kinetic roughening phenomena, stochastic growth, directed polymers and all that. Aspects of multidisciplinary statistical mechanics},
  author={Halpin-Healy, Timothy and Zhang, Yi-Cheng},
  journal={Phys. Rep.},
  volume={254},
  number={4-6},
  pages={215--414},
  year={1995},
  publisher={Elsevier},
  doi={10.1016/0370-1573(94)00087-J}
}

@misc{mcculloch2026bayesian,
  title={Bayesian Tracking of a Diffusing Target in Two and Three Dimensions},
  author={McCulloch, Ewan and Nahum, Adam},
  journal={arXiv preprint arXiv:2609.00144},
  year={2026},
  url={https://arxiv.org/pdf/2609.00144}
}

@book{comets2017directed,
  title={Directed polymers in random environments},
  author={Comets, Francis and others},
  year={2017},
  publisher={Springer}
}

@article{PhysRevLett.129.260603,
	title        = {Kardar-Parisi-Zhang Physics and Phase Transition in a Classical Single Random Walker under Continuous Measurement},
	author       = {Jin, Tony and Martin, David G.},
	year         = 2022,
	month        = {Dec},
	journal      = {Phys. Rev. Lett.},
	publisher    = {American Physical Society},
	volume       = 129,
	pages        = 260603,
	doi          = {10.1103/PhysRevLett.129.260603},
	url          = {https://link.aps.org/doi/10.1103/PhysRevLett.129.260603},
	issue        = 26,
	numpages     = 6
}

@article{kim2024planteddirectedpolymerinferring,
  title={Planted directed polymer: Inferring a random walk from noisy images},
  author={P. Kim, Sun Woo and Lamacraft, Austen},
  journal={Phys. Rev. E},
  volume={111},
  number={2},
  pages={024135},
  year={2025},
  publisher={APS},
  url={https://journals.aps.org/pre/pdf/10.1103/PhysRevE.111.024135}
}

@book{kamenev2023field,
	title        = {Field theory of non-equilibrium systems},
	author       = {Kamenev, Alex},
	year         = 2023,
	publisher    = {Cambridge University Press}
}

@book{fradkin2013field,
	title        = {Field theories of condensed matter physics},
	author       = {Fradkin, Eduardo},
	year         = 2013,
	publisher    = {Cambridge University Press}
}

@article{PhysRevX.9.031009,
	title        = {Measurement-Induced Phase Transitions in the Dynamics of Entanglement},
	author       = {Skinner, Brian and Ruhman, Jonathan and Nahum, Adam},
	year         = 2019,
	month        = {Jul},
	journal      = {Phys. Rev. X},
	publisher    = {American Physical Society},
	volume       = 9,
	pages        = {031009},
	doi          = {10.1103/PhysRevX.9.031009},
	url          = {https://link.aps.org/doi/10.1103/PhysRevX.9.031009},
	issue        = 3,
	numpages     = 21
}

@article{li2018quantum,
	title = {Quantum Zeno effect and the many-body entanglement transition},
	author = {Li, Yaodong and Chen, Xiao and Fisher, Matthew P. A.},
	year         = 2018,
	month        = {Nov},
	journal      = {Phys. Rev. B},
	publisher    = {American Physical Society},
	volume       = 98,
	pages        = 205136,
	doi          = {10.1103/PhysRevB.98.205136},
	url          = {https://link.aps.org/doi/10.1103/PhysRevB.98.205136},
	issue        = 20,
	numpages     = 9
}

@article{PhysRevB.106.024305,
	title        = {Measurement-induced phase transition in a chaotic classical many-body system},
	author       = {Willsher, Josef and Liu, Shu-Wei and Moessner, Roderich and Knolle, Johannes},
	year         = 2022,
	month        = {Jul},
	journal      = {Phys. Rev. B},
	publisher    = {American Physical Society},
	volume       = 106,
	pages        = {024305},
	doi          = {10.1103/PhysRevB.106.024305},
	url          = {https://link.aps.org/doi/10.1103/PhysRevB.106.024305},
	issue        = 2,
	numpages     = 7
}

@misc{giachetti2023elusivephasetransitionreplica,
      title={Elusive phase transition in the replica limit of monitored systems}, 
      author={Guido Giachetti and De Luca, Andrea},
      year={2023},
      eprint={2306.12166},
      archivePrefix={arXiv},
      primaryClass={cond-mat.stat-mech},
      url={https://arxiv.org/abs/2306.12166}, 
}

@article{turkeshi2022measurementinducedcriticality,
	title        = {Measurement-induced criticality as a data-structure transition},
	author       = {Turkeshi, Xhek},
	year         = 2022,
	month        = {Oct},
	journal      = {Phys. Rev. B},
	publisher    = {American Physical Society},
	volume       = 106,
	pages        = 144313,
	doi          = {10.1103/PhysRevB.106.144313},
	url          = {https://link.aps.org/doi/10.1103/PhysRevB.106.144313},
	issue        = 14,
	numpages     = 9
}

@article{sierant2022dissipativefloquet,
	title        = {Dissipative {F}loquet {D}ynamics: from {S}teady {S}tate to {M}easurement {I}nduced {C}riticality in {T}rapped-ion {C}hains},
	author       = {Sierant, Piotr and Chiriac{\`{o}}, Giuliano and Surace, Federica M. and Sharma, Shraddha and Turkeshi, Xhek and Dalmonte, Marcello and Fazio, Rosario and Pagano, Guido},
	year         = 2022,
	journal      = {{Quantum}},
	publisher    = {{Verein zur F{\"{o}}rderung des Open Access Publizierens in den Quantenwissenschaften}},
	volume       = 6,
	pages        = 638,
	doi          = {10.22331/q-2022-02-02-638},
	issn         = {2521-327X},
	url          = {https://doi.org/10.22331/q-2022-02-02-638}
}

@phdthesis{brunet2000influence,
  title={Influence des effets de taille finie sur la propagation d'un front \& Distribution de l'{\'e}nergie libre d'un polym{\`e}re dirig{\'e} en milieu al{\'e}atoire},
  author={Brunet, {\'E}ric},
  year={2000},
  school={Universit{\'e} Paris-Diderot-Paris VII}
}

@article{le2012kpz,
  title={The KPZ equation with flat initial condition and the directed polymer with one freeend},
  author={Le Doussal, Pierre and Calabrese, Pasquale},
  journal={J. Stat. Mech.: Theory Exp.},
  volume={2012},
  number={06},
  pages={P06001},
  year={2012},
  publisher={IOP Publishing},
  url={https://iopscience.iop.org/article/10.1088/1742-5468/2012/06/P06001}
}

@article{pizzi2022bridgingthegap,
	title        = {Bridging the gap between classical and quantum many-body information dynamics},
	author       = {Pizzi, Andrea and Malz, Daniel and Nunnenkamp, Andreas and Knolle, Johannes},
	year         = 2022,
	month        = {Dec},
	journal      = {Phys. Rev. B},
	publisher    = {American Physical Society},
	volume       = 106,
	pages        = 214303,
	doi          = {10.1103/PhysRevB.106.214303},
	url          = {https://link.aps.org/doi/10.1103/PhysRevB.106.214303},
	issue        = 21,
	numpages     = 15
}

@article{li2021statisticalmechanicsof,
	title        = {Statistical mechanics of quantum error correcting codes},
	author       = {Li, Yaodong and Fisher, Matthew P. A.},
	year         = 2021,
	month        = {Mar},
	journal      = {Phys. Rev. B},
	publisher    = {American Physical Society},
	volume       = 103,
	pages        = 104306,
	doi          = {10.1103/PhysRevB.103.104306},
	url          = {https://link.aps.org/doi/10.1103/PhysRevB.103.104306},
	issue        = 10,
	numpages     = 19
}

@article{gerbino2025measurement,
  title={Measurement-induced phase transition in state estimation of chaotic systems and the directed polymer},
  author={Gerbino, Federico and Giachetti, Guido and Le Doussal, Pierre and De Luca, Andrea},
  journal={Phys. Rev. Res.},
  volume={7},
  number={3},
  pages={033105},
  year={2025},
  publisher={APS},
  url = {https://journals.aps.org/prresearch/abstract/10.1103/6375-8ncz}
}

@article{Le_Doussal_2016,
   title={Large deviations for the height in 1D Kardar-Parisi-Zhang growth at late times},
   volume={113},
   ISSN={1286-4854},
   url={http://dx.doi.org/10.1209/0295-5075/113/60004},
   DOI={10.1209/0295-5075/113/60004},
   number={6},
   journal={Europhy. Lett.},
   publisher={IOP Publishing},
   author={Le Doussal, Pierre and Majumdar, Satya N. and Schehr, Grégory},
   year={2016},
   month=mar, pages={60004} }

@misc{wiese2026bayesian,
      title={Bayesian phase transition for the critical Ising model: Enlarged replica symmetry in the epsilon expansion and in 2D}, 
      author={Kay Joerg Wiese and Alapan Das and Adam Nahum},
      year={2026},
      eprint={2604.23346},
      archivePrefix={arXiv},
      primaryClass={cond-mat.stat-mech},
      url={https://arxiv.org/abs/2604.23346}, 
}

@article{PhysRevResearch.4.L022066,
  title = {Dynamics of measured many-body quantum chaotic systems},
  author = {Altland, Alexander and Buchhold, Michael and Diehl, Sebastian and Micklitz, Tobias},
  journal = {Phys. Rev. Res.},
  volume = {4},
  issue = {2},
  pages = {L022066},
  numpages = {6},
  year = {2022},
  month = {Jun},
  publisher = {American Physical Society},
  doi = {10.1103/PhysRevResearch.4.L022066},
  url = {https://link.aps.org/doi/10.1103/PhysRevResearch.4.L022066}
}

@article{PhysRevB.107.014306,
  title = {Entanglement structure in the volume-law phase of hybrid quantum automaton circuits},
  author = {Han, Yiqiu and Chen, Xiao},
  journal = {Phys. Rev. B},
  volume = {107},
  issue = {1},
  pages = {014306},
  numpages = {16},
  year = {2023},
  month = {Jan},
  publisher = {American Physical Society},
  doi = {10.1103/PhysRevB.107.014306},
  url = {https://link.aps.org/doi/10.1103/PhysRevB.107.014306}
}

@article{SciPostPhysCore.5.2.023,
	title = {Measurement-induced criticality in extended and long-range unitary circuits},
	pages = {023},
	author = {Sharma, Shraddha and Turkeshi, Xhek and Fazio, Rosario and Dalmonte, Marcello},
	journal = {SciPost Phys. Core},
	volume = {5},
	year = {2022},
	publisher = {SciPost},
	doi = {10.21468/SciPostPhysCore.5.2.023},
	url = {https://scipost.org/10.21468/SciPostPhysCore.5.2.023}
}

@article{PRXQuantum.6.010351,
  title = {Concomitant Entanglement and Control Criticality Driven by Collective Measurements},
  author = {Iadecola, Thomas and Wilson, Justin H. and Pixley, J.H.},
  journal = {PRX Quantum},
  volume = {6},
  issue = {1},
  pages = {010351},
  numpages = {19},
  year = {2025},
  month = {Mar},
  publisher = {American Physical Society},
  doi = {10.1103/PRXQuantum.6.010351},
  url = {https://link.aps.org/doi/10.1103/PRXQuantum.6.010351}
}

@article{skinner2019measurementinducedphase,
  title = {Measurement-Induced Phase Transitions in the Dynamics of Entanglement},
  author = {Skinner, Brian and Ruhman, Jonathan and Nahum, Adam},
  journal = {Phys. Rev. X},
  volume = {9},
  issue = {3},
  pages = {031009},
  numpages = {21},
  year = {2019},
  month = {Jul},
  publisher = {American Physical Society},
  doi = {10.1103/PhysRevX.9.031009},
  url = {https://link.aps.org/doi/10.1103/PhysRevX.9.031009}
}

@article{li2019measurementdrivenentanglement,
  title = {Measurement-driven entanglement transition in hybrid quantum circuits},
  author = {Li, Yaodong and Chen, Xiao and Fisher, Matthew P. A.},
  journal = {Phys. Rev. B},
  volume = {100},
  issue = {13},
  pages = {134306},
  numpages = {26},
  year = {2019},
  month = {Oct},
  publisher = {American Physical Society},
  doi = {10.1103/PhysRevB.100.134306},
  url = {https://link.aps.org/doi/10.1103/PhysRevB.100.134306}
}

@article{chan2019unitaryprojective,
  title = {Unitary-projective entanglement dynamics},
  author = {Chan, Amos and Nandkishore, Rahul M. and Pretko, Michael and Smith, Graeme},
  journal = {Phys. Rev. B},
  volume = {99},
  issue = {22},
  pages = {224307},
  numpages = {16},
  year = {2019},
  month = {Jun},
  publisher = {American Physical Society},
  doi = {10.1103/PhysRevB.99.224307},
  url = {https://link.aps.org/doi/10.1103/PhysRevB.99.224307}
}

@article{szyniszewski2019entanglementtransitionfrom,
  title = {Entanglement transition from variable-strength weak measurements},
  author = {Szyniszewski, M. and Romito, A. and Schomerus, H.},
  journal = {Phys. Rev. B},
  volume = {100},
  issue = {6},
  pages = {064204},
  numpages = {8},
  year = {2019},
  month = {Aug},
  publisher = {American Physical Society},
  doi = {10.1103/PhysRevB.100.064204},
  url = {https://link.aps.org/doi/10.1103/PhysRevB.100.064204}
}

@inbook{potter2022entanglementdynamicsin,
    author="Potter, Andrew C. and Vasseur, Romain",
    editor="Bayat, Abolfazl and Bose, Sougato and Johannesson, Henrik",
    title="Entanglement Dynamics in Hybrid Quantum Circuits",
    bookTitle="Entanglement in Spin Chains: From Theory to Quantum Technology Applications",
    year="2022",
    publisher="Springer",
    address="Cham",
    pages="211--249",
    doi="10.1007/978-3-031-03998-0_9",
    url="https://doi.org/10.1007/978-3-031-03998-0_9"
}

@Inbook{lunt2022quantumsimulationusing,
author="Lunt, Oliver
and Richter, Jonas
and Pal, Arijeet",
editor="Bayat, Abolfazl
and Bose, Sougato
and Johannesson, Henrik",
title="Quantum Simulation Using Noisy Unitary Circuits and Measurements",
bookTitle="Entanglement in Spin Chains: From Theory to Quantum Technology Applications",
year="2022",
publisher="Springer",
address="Cham",
pages="251--284",
isbn="978-3-031-03998-0",
doi="10.1007/978-3-031-03998-0_10",
url="https://doi.org/10.1007/978-3-031-03998-0_10"
}

@article{nahum2017quantum,
	title = {Quantum Entanglement Growth under Random Unitary Dynamics},
	author = {Nahum, Adam and Ruhman, Jonathan and Vijay, Sagar and Haah, Jeongwan},
	journal = {Phys. Rev. X},
	volume = {7},
	issue = {3},
	pages = {031016},
	numpages = {30},
	year = {2017},
	month = {Jul},
	publisher = {American Physical Society},
	doi = {10.1103/PhysRevX.7.031016},
	url = {https://link.aps.org/doi/10.1103/PhysRevX.7.031016}
}

@article{nahum2018operator,
	title = {Operator Spreading in Random Unitary Circuits},
	author = {Nahum, Adam and Vijay, Sagar and Haah, Jeongwan},
	journal = {Phys. Rev. X},
	volume = {8},
	issue = {2},
	pages = {021014},
	numpages = {30},
	year = {2018},
	month = {Apr},
	publisher = {American Physical Society},
	doi = {10.1103/PhysRevX.8.021014},
	url = {https://link.aps.org/doi/10.1103/PhysRevX.8.021014}
}

@article{von_Keyserlingk_2018,
  title = {Operator Hydrodynamics, OTOCs, and Entanglement Growth in Systems without Conservation Laws},
  author = {von Keyserlingk, C. W. and Rakovszky, Tibor and Pollmann, Frank and Sondhi, S. L.},
  journal = {Phys. Rev. X},
  volume = {8},
  issue = {2},
  pages = {021013},
  numpages = {19},
  year = {2018},
  month = {Apr},
  publisher = {American Physical Society},
  doi = {10.1103/PhysRevX.8.021013},
  url = {https://link.aps.org/doi/10.1103/PhysRevX.8.021013}
}

@article{zhang2021emergentreplica,
  doi = {10.22331/q-2021-11-16-579},
  url = {https://doi.org/10.22331/q-2021-11-16-579},
  title = {Emergent {R}eplica {C}onformal {S}ymmetry in {N}on-{H}ermitian {SYK}{$_2$} {C}hains},
  author = {Zhang, Pengfei and Jian, Shao-Kai and Liu, Chunxiao and Chen, Xiao},
  journal = {{Quantum}},
  issn = {2521-327X},
  publisher = {{Verein zur F{\"{o}}rderung des Open Access Publizierens in den Quantenwissenschaften}},
  volume = {5},
  pages = {579},
  month = nov,
  year = {2021}
}

@article{turkeshi2020measurementinducedcriticality,
  title = {Measurement-induced criticality in $(2+1)$-dimensional hybrid quantum circuits},
  author = {Turkeshi, Xhek and Fazio, Rosario and Dalmonte, Marcello},
  journal = {Phys. Rev. B},
  volume = {102},
  issue = {1},
  pages = {014315},
  numpages = {9},
  year = {2020},
  month = {Jul},
  publisher = {American Physical Society},
  doi = {10.1103/PhysRevB.102.014315},
  url = {https://link.aps.org/doi/10.1103/PhysRevB.102.014315}
}

@article{sierant2022measurement,
  title = {Measurement-induced phase transitions in $(d+1)$-dimensional stabilizer circuits},
  author = {Sierant, Piotr and Schir\`o, Marco and Lewenstein, Maciej and Turkeshi, Xhek},
  journal = {Phys. Rev. B},
  volume = {106},
  issue = {21},
  pages = {214316},
  numpages = {20},
  year = {2022},
  month = {Dec},
  publisher = {American Physical Society},
  doi = {10.1103/PhysRevB.106.214316},
  url = {https://link.aps.org/doi/10.1103/PhysRevB.106.214316}
}

@article{sierant2022universalbehaviorbeyond,
  title = {Universal Behavior beyond Multifractality of Wave Functions at Measurement-Induced Phase Transitions},
  author = {Sierant, Piotr and Turkeshi, Xhek},
  journal = {Phys. Rev. Lett.},
  volume = {128},
  issue = {13},
  pages = {130605},
  numpages = {7},
  year = {2022},
  month = {Apr},
  publisher = {American Physical Society},
  doi = {10.1103/PhysRevLett.128.130605},
  url = {https://link.aps.org/doi/10.1103/PhysRevLett.128.130605}
}

@article{Noel_2022,
   title={Measurement-induced quantum phases realized in a trapped-ion quantum computer},
   volume={18},
   ISSN={1745-2481},
   url={http://dx.doi.org/10.1038/s41567-022-01619-7},
   DOI={10.1038/s41567-022-01619-7},
   number={7},
   journal={Nat. Phys.},
   publisher={Springer Science and Business Media LLC},
   author={Noel, Crystal and Niroula, Pradeep and Zhu, Daiwei and Risinger, Andrew and Egan, Laird and Biswas, Debopriyo and Cetina, Marko and Gorshkov, Alexey V. and Gullans, Michael J. and Huse, David A. and Monroe, Christopher},
   year={2022},
   month= jun, pages={760–764} }

@article{google2021,
   title={Exponential suppression of bit or phase errors with cyclic error correction},
   volume={595},
   ISSN={1476-4687},
   url={http://dx.doi.org/10.1038/s41586-021-03588-y},
   DOI={10.1038/s41586-021-03588-y},
   number={7867},
   journal={Nature},
   publisher={Springer Science and Business Media LLC},
   author = {{Google Quantum AI}}, 
   year={2021},
   month= jul, pages={383–387} }

@Article{feng2025postselectionfreeexperimental,
author={Feng, Xiaozhou
and C{\^o}t{\'e}, Jeremy
and Kourtis, Stefanos
and Skinner, Brian},
title={Postselection-free experimental observation of the measurement-induced phase transition in circuits with universal gates},
journal={Commun. Phys.},
year={2026},
month={Feb},
day={14},
volume={9},
number={1},
pages={110},
issn={2399-3650},
doi={10.1038/s42005-025-02443-0},
url={https://doi.org/10.1038/s42005-025-02443-0}
}

@article{koh2022experimentalrealizationof,
  title={Measurement-induced entanglement phase transition on a superconducting quantum processor with mid-circuit readout},
  author={Koh, Jin Ming and Sun, Shi-Ning and Motta, Mario and Minnich, Austin J},
  journal={Nat. Phys.},
  volume={19},
  number={9},
  pages={1314--1319},
  year={2023},
  url = {https://www.nature.com/articles/s41567-023-02076-6},
  publisher={Nature Publishing Group UK London}
}

@article{gerbino2024dyson,
  title={A Dyson Brownian motion model for weak measurements in chaotic quantum systems},
  author={Gerbino, Federico and Le Doussal, Pierre and Giachetti, Guido and {De Luca}, Andrea},
  journal={Quantum Rep.},
  volume={6},
  number={2},
  pages={200--230},
  year={2024},
  publisher={MDPI},
  url = {https://www.mdpi.com/2624-960X/6/2/16}
}

@article{deluca2025universality,
  title = {Universality Classes for Purification in Nonunitary Quantum Processes},
  author = {De Luca, Andrea and Liu, Chunxiao and Nahum, Adam and Zhou, Tianci},
  date = {2025-11-07},
  year={2025},
  journal={Phys. Rev. X},
  journaltitle = {Phys. Rev. X},
  shortjournal = {Phys. Rev. X},
  volume = {15},
  number = {4},
  pages = {041024},
  issn = {2160-3308},
  doi = {10.1103/wlj6-mkk4},
  url = {https://link.aps.org/doi/10.1103/wlj6-mkk4},
  urldate = {2026-02-12}
}

@article{sierant2023,
  title = {Entanglement Growth and Minimal Membranes in ($d+1$) Random Unitary Circuits},
  author = {Sierant, Piotr and Schir\`o, Marco and Lewenstein, Maciej and Turkeshi, Xhek},
  journal = {Phys. Rev. Lett.},
  volume = {131},
  issue = {23},
  pages = {230403},
  numpages = {7},
  year = {2023},
  month = {Dec},
  publisher = {American Physical Society},
  doi = {10.1103/PhysRevLett.131.230403},
  url = {https://link.aps.org/doi/10.1103/PhysRevLett.131.230403}
}

@article{nahum2025bayesian,
  title = {Bayesian critical points in classical lattice models},
  author = {Nahum, Adam and Jacobsen, Jesper Lykke},
  journal = {Phys. Rev. B},
  volume = {112},
  issue = {23},
  pages = {235113},
  numpages = {57},
  year = {2025},
  month = {Dec},
  publisher = {American Physical Society},
  doi = {10.1103/7dpt-d4s5},
  url = {https://link.aps.org/doi/10.1103/7dpt-d4s5}
}

@article{ippoliti2021entanglement,
  title = {Entanglement Phase Transitions in Measurement-Only Dynamics},
  author = {Ippoliti, Matteo and Gullans, Michael J. and Gopalakrishnan, Sarang and Huse, David A. and Khemani, Vedika},
  journal = {Phys. Rev. X},
  volume = {11},
  issue = {1},
  pages = {011030},
  numpages = {23},
  year = {2021},
  month = {Feb},
  publisher = {American Physical Society},
  doi = {10.1103/PhysRevX.11.011030},
  url = {https://link.aps.org/doi/10.1103/PhysRevX.11.011030}
}

@article{block2022measurement,
  title = {Measurement-Induced Transition in Long-Range Interacting Quantum Circuits},
  author = {Block, Maxwell and Bao, Yimu and Choi, Soonwon and Altman, Ehud and Yao, Norman Y.},
  journal = {Phys. Rev. Lett.},
  volume = {128},
  issue = {1},
  pages = {010604},
  numpages = {7},
  year = {2022},
  month = {Jan},
  publisher = {American Physical Society},
  doi = {10.1103/PhysRevLett.128.010604},
  url = {https://link.aps.org/doi/10.1103/PhysRevLett.128.010604}
}

\clearpage

\section{End Matter}

\textit{A) Continuum limit.}---Consider the diffusive scaling $x=j \Delta x$, $t=\t \Delta t$, with time-step and lattice spacing related by $\Delta t=\Delta x^2/2 \to 0$. Since the walk occupies at each time only one of the two sublattices, the sites it can reach are spaced by $2\Delta x$; we therefore introduce the density $\z_{\Delta t}(x,t) := z_{x/\Delta x}^{(t/\Delta t)}/(2\Delta x)$, with $2\Delta x = 2\sqrt{2\Delta t}$, for which Eq.~\eqref{eq:discreteSHE} reads
\begin{equation} \label{eq:SHEem}
\small 
    \partial_t \z_{\Delta t} = \partial^2_x \z_{\Delta t} + \frac{1}{\Delta t}  \eta_{x/\Delta x}^{(t/\Delta t)}\ \z_{\Delta t} \,,\quad
    \eta_j^{(\t)} := \frac{P_{1,\t}(a_j^{(\t)})}{P_{0,\t}(a_j^{(\t)})}-1 \,,
\end{equation}
with $\average{\eta_j^{(\t)}}_0=0$. A nontrivial $\Delta t\to0$ limit exists only if the two distributions become indistinguishable, according to the weak-noise scaling ($a$-dependence implied)
\begin{equation}\label{eq:scaling}
    P_{0,\t} = P_{1,\t} + \pi_{1,\t} \Delta t^{1/4} + \pi_{2,\t} \Delta t^{1/2} + O(\Delta t^{3/4}) \,,
\end{equation}
with $\int da \ \pi_{1/2,\t}(a) = 0$ fixed by normalization. Then $\eta_j^{(\t)}=O(\Delta t^{1/4})$ and the variance
\begin{equation} \label{eq:cdef}
    c(t=\t \Delta t) := \sqrt{\frac{2}{\Delta t}}\average{(\eta_j^{(\t)})^2}_0 \simeq \sqrt 2\int_{\mathbb R} da \frac{\pi_{1,\t}^2(a)}{P_{1,\t}(a)}
\end{equation}
is finite, $\Delta t^{-1}\eta^{(t/\Delta t)}_{x/\Delta x}\to \sqrt{2c(t)}\,\xi(x,t)$ with $\xi$ a standard white noise, and $\z_{\Delta t}$ converges to the solution $z(x,t)$ of Eq.~\eqref{eq:SHE}. Since $\average{(\eta_j^{(\t)})^2}_0=\int da\,(P_{1,\t}-P_{0,\t})^2/P_{0,\t}\simeq 2D_{\rm KL}(P_{1,\t}||P_{0,\t})$, the coupling $c(t)$ is fixed by the information gained in a single round of measurements, $c(t)=2\sqrt{2/\Delta t}\ D_{\rm KL}(P_{1,\t}||P_{0,\t})$~\cite{gerbino2024dyson}.

In these variables the Shannon entropy \eqref{eq:entropydef} reads
\begin{equation} \label{eq:Sdiv}
\begin{split}
    \average{S_\t} = & -\ln {2\Delta x}+
    \average{ {\mathrm Z}_{\Delta t}(t)\ln {\mathrm Z}_{\Delta t}(t)}_0 +\\
    & - \sum_x 2\Delta x\  \average{\z_{\Delta t}(x,t)\ln \z_{\Delta t}(x,t)}_0  \,,
\end{split}
\end{equation}
with ${\mathrm Z}_{\Delta t}(t):=\sum_x 2\Delta x\ \z_{\Delta t}(x,t)$ and the sum running over the sites reached by the walk. The last two terms converge respectively to $S_{\rm F}$ and $S_{\rm D}$ of Eq.~\eqref{eq:replicamoments}, while the diverging $-\ln {2\Delta x}$ is a pure lattice artifact, subtracted throughout to match the continuum convention \eqref{eq:replicaentropy}.

\textit{B) Expansion around free diffusion: $\alpha>1/2$.}---The two regimes of Eq.~\eqref{eq:mainresult} are governed by the flow of the dimensionless coupling $g(t)=c(t)\sqrt t$ of Eq.~\eqref{eq:gdimless}, and require two different expansions. For $\alpha>1/2$ the replicas are asymptotically free and we expand in the LL interaction. In the interaction picture with respect to $\hat H_0=-\sum_i\nabla_i^2$, i.e., $\ket{\Psi_I(t)}:=e^{\hat H_0(t-t_0)}\ket{\Psi(t)}$, Eq.~\eqref{eq:schrodinger} gives
\begin{equation} \label{eq:dyson}
\begin{split}
    &\ket{\Psi_I(t)} = \mathcal T e^{-\int_{t_0}^{t} dt'\, \hat V_I(t')}\ket{\Psi(t_0)}\,, \\
    &\hat V_I(t) = -2c(t)\, e^{\hat H_0(t-t_0)}\sum_{i<j}\delta(x_i-x_j)\, e^{-\hat H_0(t-t_0)} \,,
\end{split}
\end{equation}
which we truncate at first order in $c_0$. Using the free propagator $G_n(\xvec;t) = e^{-\sum_i x_i^2/(4t)}/(4\pi t)^{n/2}$ and the approximation $\langle \yvec\ket{\Psi(t_0)}\approx \prod_i \delta(y_i)$ discussed in the main text, the two moments \eqref{eq:replicamoments} read
\begin{equation} \label{eq:pertmoments}
\begin{split}
    & \overbar{Z(t)^n} = \left[1+n(n-1)\,c_0\, C_{\rm F}(t)\right]+O(c_0^2) \,,\\
    & \int_{\mathbb R} dx\ \overbar{z(x,t)^n} = \frac{\left[1+n(n-1)\,c_0\, C_{\rm D}(t)\right]}{\sqrt n\ [4\pi (t-t_0)]^{\frac{n-1}{2}}}+O(c_0^2)\,,
\end{split}
\end{equation}
with
\begin{equation} \label{eq:CFCD}
\begin{split}
    & C_{\rm F}(t) = \frac{t_0^\alpha}{\sqrt{8\pi}}\int_{t_0}^{t} \frac{dt'\ t'^{-\alpha}}{\sqrt{t'-t_0}}\,,\\
    & C_{\rm D}(t) = \frac{t_0^\alpha}{\sqrt{8\pi}}\int_{t_0}^{t} \frac{dt'\ t'^{-\alpha}}{\sqrt{t'-t_0}}\ \sqrt{\frac{t-t_0}{t-t'}}\,.
\end{split}
\end{equation}
Both have a transparent interpretation: $c(t')\,[8\pi(t'-t_0)]^{-1/2}dt'$ is the monitoring-weighted probability that a given pair of replicas meets in $[t',t'+dt']$; in the droplet geometry the pair is further conditioned to meet again at time $t$, which enhances the coincidence probability by the Brownian-bridge factor $\sqrt{(t-t_0)/(t-t')}$, whence $C_{\rm D}>C_{\rm F}$. Using $\partial_n[n(n-1)]|_{n=1}=1$, Eq.~\eqref{eq:replicaentropy} gives
\begin{equation} \label{eq:Spert}
    \s[t] = \tfrac12\big[1+\ln 4\pi(t-t_0)\big] - c_0 \big[C_{\rm D}-C_{\rm F}\big] + O(c_0^2)\,,
\end{equation}
i.e., Eq.~\eqref{eq:firstterm}. Note that $C_{\rm F}$ and $C_{\rm D}$ separately converge, for $\alpha>1/2$, to the \emph{same} constant, dominated by $t'\approx t_0$ where the expansion is not controlled; only their difference, dominated by $t'\sim t$, enters the entropy. Setting $t'=tu$, one may let $t_0/t\to0$, and
\begin{equation} \label{eq:Ialpha}
\begin{split}
    & C_{\rm D}-C_{\rm F} \simeq \frac{t_0^\alpha}{\sqrt{8\pi}}\, I(\alpha)\ t^{\frac12-\alpha}\,,\\
    & I(\alpha)=\int_0^1 du\ u^{-\alpha-\frac12}\big[(1-u)^{-\frac12}-1\big]\,,
\end{split}
\end{equation}
which is positive and finite for $1/2<\alpha<3/2$ and yields 
\begin{equation} \label{eq:Ialpha_min}
    I(\alpha) = \sqrt{\pi} \ \frac{\Gamma(\tfrac12-\alpha)}{\Gamma(1-\alpha)}+\frac{1}{\alpha-1/2}>0 \,.
\end{equation}
For $\alpha \to 1/2$, $I(\alpha)$ has a smooth limit $\lim_{\alpha\to 1/2^+}I(\alpha)=2\ln2$, since the two poles in Eq.~\eqref{eq:Ialpha_min} cancel as $\alpha\to1/2^+$. The relative size of the correction is $O(c_0t_0^\alpha t^{1/2-\alpha})=O(g(t))$, so the expansion is self-consistent at long times for every $\alpha>1/2$. We refer to~\cite{supplnote} for additional details.

\textit{C) Expansion around the Lieb-Liniger ground state: $0<\alpha<1/2$.}---Here $g(t)\propto t^{1/2 - \alpha}\to\infty$ and the previous expansion fails. The change of variables of Ref.~\cite{barraquand2020stochastic} used in the main text,
\be
\begin{aligned}
& y=c(t)\ x,\quad  \tau(t)=\int_0^t d s\, c(s)^2,\\
& z(x,t)=\sqrt{c(t)}\,
\exp\!\left[\frac{c'(\tau)}{4c(\tau)}y^2\right]\zz(y,\tau),
\end{aligned}
\label{eq:mapping}
\ee
trades the decaying coupling for the fixed coupling $\bar c=1$ plus the harmonic trap of Eq.~\eqref{eq:SHEharmonic}. For the protocol~\eqref{eq:ct},
\begin{equation} \label{eq:thetadef}
\small
    \tau(t) = \tau_0+\frac{c_0^2 t_0}{1-2\alpha}\left( \frac{t}{t_0} \right)^{1-2\alpha},\
    \tau_0 = \int_0^{t_0}\!\!\! ds\, c(s)^2 - \frac{c_0^2 t_0}{1-2\alpha},
\end{equation}
with $\tau_0$ collecting all the short-time information on $c(t)$, and $A(\tau)$ is given by Eq.~\eqref{eq:potentialdef}. The moments \eqref{eq:replicamoments} become
\begin{subequations} \label{eq:replicamapped}
\begin{equation}
    \overline{Z(t)^n} = \int_{\mathbb R^n} \frac
    {d^ny}{c(\tau)^{n/2}}\ e^{\frac{c'(\tau)}{4c(\tau)}\sum_{i} y_i^2} \ \overbar{\prod\nolimits_{j} \zz(y_j,\tau)} \,,
    \label{eq:ZTn}
\end{equation}
\begin{equation}
\label{eq:ztn}
     \int_{\mathbb R} dx\ \overline{z(x,t)^n} = c(\tau)^{\frac{n}{2}-1} \int_{\mathbb R} dy \ e^{\frac{c'(\tau)}{4c(\tau)} n y^2} \overline{\zz(y,\tau)^n} \,.
\end{equation}
\end{subequations}
In Eq.~\eqref{eq:ZTn}, we change variables $d^n y \to dR\, d^{n-1} \tilde y$ and we use the decomposition~\eqref{eq:COMREL} rewriting the exponent as $\sum_i y_i^2 = nR^2+\f(\tyvec)$. 
The integration over $R$ can be performed using that 
the center-of-mass kernel solves 
\be
\partial_\tau K_n(R,\tau)
= \frac{1}{n}\partial_R^2 K_n(R,\tau)
-\frac{n}{2}A(\tau) R^2\,K_n(R,\tau) \,,
\label{eq:COMeq}
\ee
with initial condition $K_n(R,0) =\delta(R)$. 
Without the explicit form of the kernel $K_n(R, \tau)$, we can use that, for \emph{any} profile $c(\tau)$ and $A=c''/(2c)$, the weight $W(R,\tau):= c(\tau)^{-1/2}e^{n c'R^2/(4c)}$ obeys the adjoint equation $\partial_\tau W = -\tfrac1n \partial^2_RW+\tfrac n2 A(\tau)R^2W$, so that $\int dR\, W K_n$ is $\tau$-independent and equal to its value at $\tau=0$, which is $1$ because $c(0)=1$. This leads to the conservation law
\begin{equation}
    \int_{\mathbb R} \frac{d R}{\sqrt{c(\tau)}} \ e^{n\frac{c'(\tau)}{4c(\tau)}R^2} K_n(R,\tau) = 1\,,
\label{eq:COM}
\end{equation}
which can then be used to obtain Eq.~\eqref{eq:ZnTmain}. For Eq.~\eqref{eq:ztn}, we proceed similarly, noting that the integration over the coinciding coordinates $y$ reduces to integration over $R$ which can be performed once again via \eqref{eq:COM} and leads to Eq.~\eqref{eq:zntmain}.

Inserting Eq.~\eqref{eq:GSproj} in 
\eqref{eq:Sdiffder}, and writing $C_n(\tau) = \kappa_n(\tau) e^{-\tau E_n^{\rm GS}}$ one has
\begin{multline}
\label{eq:entroexplicit}
    \s[t] =   - \ln c(\tau) +\\ 
    \partial_{n=1}\Bigl[ C_n(\tau) \bigl(\int
    d^{n-1}\tilde y \
e^{\frac{c'(\tau )}{4 c(\tau )}\f(\tilde \yvec)}
    \psi_n(\tilde \yvec) - \psi_n(0)\bigr)\Bigr]
    \;.
\end{multline}
Now, if we differentiate the prefactor $C_n(\tau)$ with respect to $n$, both the integral and $\psi_n(0)$ are computed for $n \to 1$, where they both give $1$. So, as mentioned in the main text, the prefactor has to be computed precisely at $n=1$ and one simply uses $\lim_{n\to1}  C_n(\tau) = 1$.

Finally, to compute the integral at large $\tau$, we expand the exponent as explained in the main text, and use the form valid in the sector $\tilde y_\ell>0$
\begin{equation} \label{eq:psin}
    \psi_n(\tyvec) = \frac{n!}{n^2}  \ e^{-\frac{1}{2} \sum_{\ell=1}^{n-1} \ell(n-\ell) \tilde y_\ell} \,,
\end{equation}
(the other sectors following by permutation symmetry) and the identities
\begin{subequations} \label{eq:psimoments}
\begin{align}\label{eq:H2}
    & \int_{\mathbb R^{n-1}} \!\!\! d^{n-1}\tilde y\ \psi_n(\tyvec) = 2^{n-1} \,,\\ 
    & \int_{\mathbb R^{n-1}} \!\!\!  d^{n-1}\tilde y\ \f(\tyvec)\,\psi_n(\tyvec) = \frac{2^{n+2}}{n}H^{(2)}_{n-1}\,,
\end{align}
\end{subequations}
with the generalized harmonic numbers $H^{(2)}_{n-1}=\sum_{k=1}^{n-1}k^{-2}=\zeta(2)-\psi'(n)$ ($\psi$ being the digamma function here). Collecting all terms, we arrive at
\begin{multline}
    \s[t] = -\ln c(\tau) + \\ 
    \partial_n\left(2^{n-1}-\frac{n!}{n^2}+\frac{2^{n+2} H_{n-1}^{(2)} c'(\tau )}{4 n  c(\tau )}\right) + O(\tau^{-2})\;.
\end{multline}
Finally, using that $\partial_nH^{(2)}_{n-1}|_{n=1}=-\psi''(1)=2\zeta(3)$ one retrieves Eq.~\eqref{eq:alphaLT}.

%%%%%%%%%%%%%%%%%%%%%%%%%%%%%%%%%%%%%%%%%%%%%%%%%%%%%%%%%%%%%%%%%%%%%%%%%%%%%%%%%%%%%%%%%%%%%%%%%%%%%%%%%%%%%%%%%%

\clearpage
\onecolumngrid

\setcounter{page}{1}
\setcounter{equation}{0}
\setcounter{figure}{0}
\setcounter{table}{0}

\setcounter{secnumdepth}{3}

\renewcommand{\thesection}{\Roman{section}}
\renewcommand{\thesubsection}{\thesection.\arabic{subsection}}
\renewcommand{\thesubsubsection}{\thesubsection.\arabic{subsubsection}}

\renewcommand{\theequation}{S\arabic{equation}}
\renewcommand{\thefigure}{S\arabic{figure}}
\renewcommand{\thetable}{S\arabic{table}}
\renewcommand{\theHequation}{S\arabic{equation}}
\renewcommand{\theHfigure}{S\arabic{figure}}
\renewcommand{\theHtable}{S\arabic{table}}

\makeatletter
\def\@sectioncntformat#1{\csname the#1\endcsname.\quad}
\makeatother

\begin{center}
    {\large \bfseries Supplemental Material:\\[0.4em]
    Bayesian Monitoring of a Diffusive Particle in One Dimension\par}
    \vspace{1em}
    {\normalsize
    Federico Gerbino$^{1}$, Guido Giachetti$^{2}$, Pierre Le Doussal$^{2}$, and Andrea De Luca$^{2}$\par}
    \vspace{0.5em}
    {\small
    $^{1}$Laboratoire de Physique Th\'eorique et Mod\`eles Statistiques, Universit\'e Paris-Saclay, CNRS, 91405 Orsay, France\\ 
    $^{2}$Laboratoire de Physique de l'\'Ecole Normale Sup\'erieure, CNRS, ENS \& PSL University, Sorbonne Universit\'e, Universit\'e Paris Cit\'e, 75005 Paris, France\par}
    \vspace{1em}
\end{center}

This Supplemental Material collects additional technical points and the derivations of the results presented in the main text. In Sec.~\ref{sec:constant} we derive the constant-monitoring result, i.e., both the saturation value $S_\infty-\ln\bar c$ of the Shannon entropy and the asymptotic approach $\delta \mathcal S$ to it, using exact KPZ results. In Sec.~\ref{sec:alphamax} we give the details of the perturbative computation for power-law monitoring at $\alpha>1/2$, sketched in the End Matter, including the marginal case $\alpha=3/2$ and the regime $\alpha>3/2$. In Sec.~\ref{sec:LLperturb} we complement the $0<\alpha<1/2$ computation of the End Matter: we give the explicit center-of-mass kernel and we use perturbation theory to compute the corrections to the Lieb-Liniger propagator in a weakening harmonic potential, showing that they do not affect the replica limit. Sec.~\ref{sec:marginal} collects what can be said about the marginal case $\alpha=1/2$. Finally, in Sec.~\ref{sec:numerics} we discuss the numerical protocol used to simulate the discrete-time dynamics of the posterior.

\section{Constant Monitoring} \label{sec:constant}

In this Section we derive the constant-rate result Eq.~\eqref{eq:alphazero} of the main text, i.e., both the saturation constant $S_\infty$ and the function $\delta \mathcal S$ controlling the approach to it. The computation is based on exact KPZ results~\cite{sasamoto2010one,Le_Doussal_2016,le2012kpz,calabrese2010free}. Thanks to the identity $\overline{S_{\bar c}(t)} = \overline{S_{\bar c=1}(\bar c^2 t)} - \ln \bar c$, we set $\bar c=1$ throughout.
We recall from the main text that $\s[t] = S_{\rm F}(t)-S_{\rm D}(t)$, that the two terms are expressed through the generating functions $g_{\rm F/D}$ by Eq.~\eqref{eq:fromZtoG}, and that the droplet term is obtained from the rescaled point-to-point partition sum $\hat z(t) = \sqrt{4\pi t}\, z(0,t)$. Indeed, using the statistical tilt symmetry identity  
\begin{equation} \label{eq:sts}
    z(x,t) \stackrel{\rm in \ law}{=} e^{-\frac{x^2}{4t}} \frac{1}{\sqrt{4\pi t}} \hat{z}(t) \,,
\end{equation}
one recovers the result Eq.~\eqref{eq:SDhat}, i.e., $S_{\rm D}(t) = \overline{\hat z(t)\ln \hat z(t)}-\frac12[1+\ln 4\pi t]$.

\subsection{Droplet contribution $S_{\rm D}(t)$}

\subsubsection{Leading order}

Let us start from the Fredholm-determinant representation Eq.~\eqref{eq:gendroplet} of the droplet generating function $g_{\rm D}$. 
The definition used here corresponds to the one of Eq.~(10) in Ref.~\cite{Le_Doussal_2016} upon the identification 
$g_{\rm D}(s,t) \equiv g_t(s t^{-1/3})$ therein. 
We expand $g_{\rm D}$ at long times, keeping the trace term only,
\begin{equation}
    g_{\rm D}(s, t) \simeq 1 -  \int_{st^{-1/3}}^\infty dr \int_{-\infty}^\infty du \frac{{\rm{Ai}}(r+u)^2}{1+e^{-t^{1/3}u}} +\dots\,,
\end{equation}
so that
\begin{equation} \label{eq:FD_exp_deriv_lead}
    \partial_s g_{\rm D}(s,t) \simeq   t^{-1/3}\int_{-\infty}^\infty du \frac{{\rm{Ai}}(st^{-1/3}+u)^2}{1+e^{-t^{1/3}u}} + \dots \,.
\end{equation}
Integrating by parts the upper line of Eq.~\eqref{eq:fromZtoG} in the main text, which we report here for clarity
\begin{equation}
\overline{\hat{z}(t) \ln \hat{z}(t)} =\int_{\mathbb R} ds\ e^s\ \left[g_{\rm D}(s+\theta(t),t) -e^{-e^{-s}}\right] \,,
\end{equation}
the boundary terms cancel and one has
\begin{equation} \label{eq:zlnzbyparts}
    \overline{\hat z(t)\ln \hat z(t)} = \int_{\mathbb R} ds \left[ e^{-e^{-s}}
    - e^s \partial_s  g_{\rm D}\Big(s + \frac{t}{12} - \frac12\ln4\pi t,\, t\Big) \right]\,.
\end{equation}
The two terms in Eq.~\eqref{eq:zlnzbyparts} are separately divergent (as $s\to+\infty$ both approach a constant), while their difference is finite. To evaluate them we introduce a convergence factor $e^{-\mu s}$, with $0<\mu<1$, compute each term at fixed $\mu$, and take $\mu\to0^+$ at the end,
\begin{equation} \label{eq:mureg}
    \overline{\hat z(t)\ln \hat z(t)} = \lim_{\mu\to0^+}\int_{\mathbb R} ds\, e^{-\mu s} \left[ e^{-e^{-s}}
    - e^s \partial_s  g_{\rm D}\Big(s + \tfrac{t}{12} - \tfrac12\ln4\pi t,\, t\Big) \right] \equiv \lim_{\mu\to0^+}\big[I_\mu^{\rm D} - \II_\mu^{\rm D}\big]+ \ldots\,,
\end{equation}
where $\ldots$ indicates higher orders in the Fredholm determinant. 
The first term in the squared brackets is elementary: with $x=e^{-s}$ one finds
\begin{equation} \label{eq:Imu}
    I_\mu^{\rm D} := \int_{\mathbb R} ds\, e^{-\mu s}\, e^{-e^{-s}} = \int_0^\infty dx\, x^{\mu-1} e^{-x} = \Gamma(\mu) = \frac1\mu - \gamma_{\rm E} + O(\mu)\,.
\end{equation}
For the second term we insert Eq.~\eqref{eq:FD_exp_deriv_lead}, exchange the order of integration, and in the $s$-integral substitute $y = (s+\tfrac{t}{12}-\tfrac12\ln4\pi t)\,t^{-1/3}+u$, so that $e^{(1-\mu)s} = (4\pi t)^{\frac{1-\mu}{2}} e^{-(1-\mu)t/12}\, e^{(1-\mu)t^{1/3}(y-u)}$. The $y$-integral is then performed with the Airy identity
\begin{equation} \label{eq:int2}
    \int_{-\infty}^\infty dy\  {\rm{Ai}}^2(y)\ e^{y a} =
    \frac{1}{\sqrt{4\pi a}} e^{\frac{a^3}{12}} \,,
\end{equation}
taken at $a=(1-\mu)t^{1/3}$, while the remaining $u$-integral gives, with $p=t^{1/3}u$,
\begin{equation} \label{eq:uint}
    \int_{-\infty}^\infty du\, \frac{e^{-(1-\mu)t^{1/3}u}}{1+e^{-t^{1/3}u}} = t^{-1/3}\int_{-\infty}^\infty dp\, \frac{e^{\mu p}}{1+e^{p}} = t^{-1/3}\,\frac{\pi}{\sin\pi\mu}\,.
\end{equation}
Collecting the factors, the powers of $t$ and the Gaussian prefactor combine into
\begin{equation} \label{eq:IImu}
    \II_\mu^{\rm D} = \frac{(4\pi t)^{-\mu/2}}{\sqrt{1-\mu}}\, \exp\!\Big[\tfrac{t}{12}\,\mu(\mu-2)(1-\mu)\Big]\,\frac{\pi}{\sin\pi\mu}
    = \frac1\mu - \frac12\ln(4\pi t) + \frac12 - \frac t6 + O(\mu)\,,
\end{equation}
where the last equality is the small-$\mu$ expansion [note that $\tfrac{t}{12}\mu(\mu-2)(1-\mu) = -\tfrac{t\mu}{6}+O(\mu^2)$, which produces the linear term $-t/6$]. Subtracting Eq.~\eqref{eq:IImu} from Eq.~\eqref{eq:Imu}, the $1/\mu$ poles cancel and the limit $\mu\to0^+$ is finite, as it should
\begin{equation}
    \overline{\hat z(t)\ln \hat z(t)} = \frac t6 + \frac12\ln(4\pi t) - \gamma_{\rm E} - \frac12 + \dots\,,
\end{equation}
and therefore, by Eq.~\eqref{eq:SDhat},
\begin{equation} \label{eq:SDlead}
    S_{\rm D}(t) = \frac t6 - (1+\gamma_{\rm E}) + \dots \,,
\end{equation}
i.e., $c_{\rm D} = -(1+\gamma_{\rm E})$ in the notation of the main text.

\subsubsection{Subleading corrections}

We now push the expansion of the Fredholm determinant to the next order, writing
\begin{equation}\label{eq:gorders}
\begin{split}
g_{\rm D}(s, t)& =   
1 - \int_{st^{-1/3}}^{+\infty} dr\  K_t(r,r) 
+ \sum_{q \geq 2} g_{\rm D}^{(q)}(s,t) \,.
\end{split}
\end{equation}
Inserting the Airy kernel Eq.~\eqref{eq:airy}, the expansion for the second-order term $g_{\rm D}^{(2)}$ is 
\begin{equation}
    g_{\rm D}^{(2)}(s,t) = \frac{1}{4}\int_{st^{-1/3}}^\infty dr \int_{st^{-1/3}}^\infty dr' \int du\int du' \frac{\left[{\rm{Ai}}(r+u){\rm{Ai}}(r'+u') - {\rm{Ai}}(r'+u) {\rm{Ai}}(r+u') \right]^2}{(1+e^{-t^{1/3}u})(1+e^{-t^{1/3}u'})} \,.
\end{equation}
We also record a different representation of the Airy kernel (Eq.~(27) in Ref.~\cite{calabrese2010free}),
\begin{equation} \label{eq:newkernel}
K_t(r,r') = 2^{-2/3} \int \frac{dv}{1 + e^{- t^{1/3} v}} \int \frac{dk}{\pi} e^{ i k 2^{-1/3} (r-r') } {\rm Ai}(k^2 + 2^{-1/3} (2 v + r + r') ) \,,
\end{equation}
which will be used below.
Differentiating in $s$ Eq.~\eqref{eq:gorders} therefore gives
\begin{align} \label{eq:FD_exp_deriv}
    &\partial_s g_{\rm D}(s,t) = t^{-1/3}\Big[K_t(\sigma,\sigma) -\mathsf B(\sigma,t)\Big]_{\sigma=st^{-1/3}}  + \dots \,,\\
    &\mathsf B(\sigma,t) := \int_\sigma^\infty dr \left[ K_t(\sigma,\sigma)K_t(r,r) - K_t(\sigma,r)^2 \right]
\end{align}
the trace piece $K_t(\sigma,\sigma)$ being the one shown in Eq.~\eqref{eq:FD_exp_deriv_lead}, which produced the leading result~\eqref{eq:SDlead}.
Inserting Eq.~\eqref{eq:FD_exp_deriv} into Eq.~\eqref{eq:zlnzbyparts} and shifting $s \mapsto s -t/12+\frac12\ln4\pi t$, the second-order contribution to $\overline{\hat z \ln\hat z}$, and hence to $S_{\rm D}$, is
\begin{equation} \label{eq:Idef}
    \mathsf{A}(t) := \int_{\mathbb R} ds\ e^s\, t^{-1/3}\mathsf B\Big(t^{-1/3}\Big(s+\frac{t}{12}-\frac12\ln 4\pi t\Big),\, t \Big)
    = e^{-t/12}\sqrt{4\pi t} \int_{\mathbb R} d\sigma\ e^{\sigma t^{1/3}}\, \mathsf B(\sigma,t) \,,
\end{equation}
so that
\begin{equation} \label{eq:SDwithI}
    S_{\rm D}(t) = \frac t6 - (1+\gamma_{\rm E})+ \mathsf{A}(t) + \ldots \,.
\end{equation}

\textbf{Integral $\mathsf{A}(t)$.} --- Since the finite-$t$ Airy kernel~\eqref{eq:airy} is positive definite, $K_t(s,s)K_t(r,r)\geq K_t(s,r)^2$ by Cauchy-Schwarz, so that $\mathsf B\geq0$ and $\mathsf{A}>0$, and the second-order droplet contribution enters $\s[t]=S_{\rm F}-S_{\rm D}$ with a negative sign.
Renaming $\sigma\to s$ and shifting $r\mapsto s+r$, the integral becomes
\begin{equation}
\begin{split}
   \mathsf{A}(t) & = e^{-t/12}\sqrt{4\pi t} \int_{\mathbb R} ds e^{s t^{1/3}}  \int_s^{+\infty} dr\ (K_t(s,s) K_t(r,r) - K_t(s,r)^2) \\
& = e^{-t/12}\sqrt{4\pi t}\int_0^{+\infty} dr  \int_{\mathbb R} ds\ e^{s t^{1/3}}   \left(K_t(s,s) K_t(r+s,r+s) - K_t(s,r+s)^2\right) \,.
\end{split}
\end{equation}
The integral in $s$ can be resolved by inserting the definition \eqref{eq:newkernel} for the kernel, and using the following Airy identity 
\begin{equation} \label{eq:int3}
    \int_{\mathbb R} dy\  {\rm{Ai}}(y+v) {\rm{Ai}}(y+w) e^{y a} = 
    \frac{1}{\sqrt{4\pi a}}  e^{\frac{a^3}{12}-\frac{a (w+v)}{2} - \frac{(w-v)^2}{4a}} \,.
\end{equation}
Explicitly, in Eq.~\eqref{eq:int3} we identify $y=2^{2/3}s$ and $a=2^{-2/3}t^{1/3}$, taking $(v,w)=(k_1^2+2^{2/3}v_1,\,k_2^2+2^{2/3}(v_2+r))$ for $K_t(s,s)K_t(s+r,s+r)$ and $(v,w)=(k_1^2+2^{2/3}v_1+2^{-1/3}r,\,k_2^2+2^{2/3}v_2+2^{-1/3}r)$ for $K_t(s,s+r)^2$; after integrating over $s$, the integral $I$ takes the form (integration over $\mathbb R$ when not specified)
\begin{equation} 
\begin{split}
\mathsf{A}(t) = e^{-t/16} & \frac{t^{1/3}}{2^{2/3}\pi} \int \frac{dv_1}{1 + e^{- t^{1/3} v_1}} \int \frac{dv_2}{1 + e^{- t^{1/3} v_2}} \int \frac{dk_1}{\sqrt{2\pi}} \int \frac{dk_2}{\sqrt{2\pi}} 
\int_0^{+\infty} dr \ 
e^{- \frac{2^{-2/3} t^{1/3}}{2} [k_1^2 + k_2^2 + 2^{2/3}(v_1+v_2)+ 2^{2/3} r]} \\
& \times \bigg[ 
e^{- \frac{1}{4 \times 2^{-2/3} t^{1/3}  } [k_2^2 - k_1^2 + 2^{2/3}(v_1-v_2) + 2^{2/3} r]^2 } 
-  e^{ i (k_1+k_2) 2^{-1/3} r } e^{- \frac{1}{4 \times 2^{-2/3} t^{1/3}  } [k_2^2 - k_1^2 + 2^{2/3}(v_1-v_2)]^2 } \bigg]  \,.
\end{split}
\end{equation}  
To continue the calculation, we first rescale $v_{1,2}\to t^{-1/3}v_{1,2}$, $k_{1,2}\to (4/t)^{1/6}k_{1,2}$, $r\to t^{-1/3}r$ and rewrite $\frac{2 e^{-v_i/2}}{1+e^{-v_i}}=\cosh^{-1}(v_i/2)$:
\begin{equation} \label{eq:Ibeta}
\begin{split}
\mathsf{A}(t) = \frac{e^{-t/16} }{2\pi t}\ \int \frac{dv_1}{\cosh(v_1/2)} \int \frac{dv_2}{\cosh(v_2/2)} 
\int & \frac{dk_1}{\sqrt{2\pi}}\ e^{-k_1^2/2}
\int \frac{dk_2}{\sqrt{2\pi}}\ e^{-k_2^2/2}
\int_0^{+\infty} \frac{dr}{2} \ e^{-r/2}\ \times \\
& \times  \left[ e^{-\frac 1 t (k_1^2-k_2^2 + v_1-v_2 + r)^2} -  e^{ i t^{-1/2} (k_1+k_2) r} e^{-\frac 1 t (k_1^2-k_2^2 + v_1-v_2)^2} \right] \,.
\end{split}
\end{equation}
Then, we can expand at large $t$ the exponentials inside the squared bracket in the second line above. Writing $\mathsf D := k_1^2-k_2^2+v_1-v_2$ and $\mu := (k_1+k_2)r$,  it expands as
\begin{equation}
\begin{split}
    \Big[\cdots\Big] = & -\frac{i\mu}{\sqrt t} - \frac1t\Big[(\mathsf D+r)^2-\mathsf D^2-\frac{\mu^2}{2}\Big] + \frac{i}{t^{3/2}}\Big[\mu \mathsf D^2+\frac{\mu^3}{6}\Big] + \frac{1}{t^2}\Big[\frac{(\mathsf D+r)^4-\mathsf D^4}{2}-\frac{\mu^2\mathsf D^2}{2}-\frac{\mu^4}{24}\Big]+O(t^{-5/2}) \,.
\end{split}
\end{equation}
The integrations in Eq.~\eqref{eq:Ibeta} can be now identified as taking the expectation over $k_{1,2}$ distributed as standard Gaussians ($\sim \mathcal{N}(0,1)$, zero mean and unit variance), $v_{1,2}$ and $r$ as random variables with normalized measures $dv_i/[2\pi\cosh(v_i/2)]$ and $e^{-r/2}dr/2$ respectively. Then, we have that the orders $t^{-1/2}$ and $t^{-3/2}$ vanish by parity in $k_{1,2}$, and the order $t^{-1}$ vanishes because  $\average{k_i^2}=1$, hence $[1-\langle (k_1+k_2)^2\rangle/2]=0$. 
At order $t^{-2}$, using $\langle r^2\rangle=8, \langle r^4\rangle=384$, $\langle v^2\rangle = \pi^2$, $\langle \mathsf D^2\rangle = 4+2\pi^2$, $\langle (k_1+k_2)^2\mathsf D^2\rangle = 24+4\pi^2$ and $\langle (k_1+k_2)^4\rangle=12$, only the term proportional to $\langle v^2 \rangle$ survives, namely
\begin{equation}
    \langle r^2\rangle\Big[3\langle \mathsf D^2\rangle - \frac{\langle (k_1+k_2)^2\mathsf D^2\rangle}{2}\Big] + \langle r^4\rangle\Big[\frac12 - \frac{\langle (k_1+k_2)^4\rangle}{24}\Big] = 32\pi^2 \,.
\end{equation}
Finally, restoring the exponential term and the normalization of the measures as in Eq.~\eqref{eq:Ibeta}, we have
\begin{equation}
    \mathsf{A}(t) = e^{-t/16}\frac{2\pi}{t} \times \frac{32\pi^2}{t^2} = e^{-t/16}\left(\frac{4\pi}{t}\right)^3>0 \,.
\end{equation}

\textbf{Total contribution.} --- Inserting the result for $\mathsf{A}(t)$ in $S_{\rm D}$ we find
\begin{equation} \label{eq:SDwithI}
    S_{\rm D}(t) = \frac t6 - (1+\gamma_{\rm E})+ e^{-t/16}\left(\frac{4\pi}{t}\right)^3 + \ldots \,.
\end{equation}

\subsection{Flat contribution $S_{\rm F}(t)$}
\label{subsec:flatpolym}

The computation of the moments of the polymer partition function $z(x,t)$ can be carried out by representing each moment as the propagator of an $n$-body LL problem. While the main text focuses on the ground state $\ket{\Psi_n^{\rm GS}}$ of the LL Hamiltonian~\eqref{eq:LL}, in which all $n$ particles sit in a single bound state, the LL spectrum is organized into states composed of $1\leq \n \leq n$ ``strings'', i.e., bound states each containing a variable number of particles.
Accordingly, the generating function $g_{\rm F}(s,t)$ for the flat geometry, Eq.~\eqref{eq:genflat}, admits the representation
\begin{equation} \label{eq:gFstrings}
    g_{\rm F}(s,t) = 1+ \sum_{\n \geq 1} \frac{\mathcal Z_{\n}(s,t)}{\n!} \,,
\end{equation}
where each $\mathcal Z_{\n}(s,t)$ acts as a canonical partition function for the $\n$-string LL problem; we refer to Ref.~\cite{le2012kpz} for its explicit computation. 
Notice that we have the identity $g_{\rm F}(s,t) \equiv g_{\lambda=(t/4)^{1/3}}\!\left(\frac{s-t/12}{(t/4)^{1/3}},\, t\right)$, where $g_{\lambda}$ refers to the original partition function defined in Eq.~(4) of~\cite{le2012kpz}; in practice, we use a ``shifted'' generating function, 
$g_{\rm F}(s,t) \equiv \tilde g_{\lambda=(t/4)^{1/3}}\!\left(\frac{s}{(t/4)^{1/3}},\, t\right)$, with $\tilde g_\lambda$ defined around Eq.~(62) of \cite{le2012kpz}. In this Subsection, we will only keep the terms $\n=1$ and $\n=2$ of the expansion \eqref{eq:gFstrings}: we will report the one-string term $\mathcal Z_1(s,t)$ in Eq.~\eqref{eq:Z1flat} (see Eq.~(63) of~\cite{le2012kpz}), and the two-string term $\mathcal Z_2(s,t)$ in Eqs.~(\ref{eq:fdef}, \ref{eq:kernel}, \ref{eq:Z2flat}) (see Eq.~(153) in Ref.~\cite{le2012kpz}). 

In order to compute $S_{\rm F}(t)$, we plug the expansion \eqref{eq:gFstrings} into the bottom line of Eq.~\eqref{eq:fromZtoG}, reported below
\begin{equation} 
 S_{\rm F}(t) = \int_{\mathbb R} ds \ e^{s}[ g_{\rm F}(s+t/12,t) - g_{\rm F}(s,0)]\,,
\end{equation}
and using $\Theta(s)$ for the Heaviside theta function, we have 
\begin{equation} \label{eq:flat_def}
\begin{split}
    S_{\rm F}(t)
    & = \int_{\mathbb R} ds\  e^s \left[1+\mathcal{Z}_1(s+\tfrac{t}{12},t) + \frac12 \mathcal{Z}_2(s+\tfrac{t}{12},t) - e^{-e^{-s}} \right]
    + \dots =  \\
    & = \underbrace{\int_{\mathbb R} ds \left[ e^s - e^{s-e^{-s}}- \Theta(s) \right]}_{1-\gamma_{\rm E}} + \underbrace{\int_{\mathbb R} ds \left[ e^s \mathcal Z_1(s+\tfrac{t}{12},t) + \Theta(s) \right]}_{A_1(t)} + \underbrace{\frac12 \int_{\mathbb R} ds\ e^s \mathcal{Z}_2(s+\tfrac{t}{12},t)}_{A_2(t)} + ...\,,
\end{split}
\end{equation}
where $\ldots$ indicates contributions from higher strings $n_s\geq 3$ in \eqref{eq:gFstrings}. 
It is convenient to write the integral as above in order to separate the constant contribution $1-\gamma_{\rm E}$ from the remaining parts $A_1(t)$ and $A_2(t)$.

\subsubsection{Leading order: the one-string term}

Let us evaluate explicitly the second integral $A_1(t)$.
The one-string term admits a closed expression, given in Eq.~(63) of~\cite{le2012kpz} with the identification $\mathcal Z_1(s,t) \equiv Z(1,s/(t/4)^{1/3})$, and reads
\begin{equation} \label{eq:Z1flat}
\begin{split}
    \mathcal Z_1(s, t) = - \frac{1}{(2t)^{1/3}}\int_{-\infty}^\infty & du \ {\rm Ai} \left(\bigl(\tfrac 4t\bigr)^{1/3}(s+u)\right)  \left[ 1-\exp(-2e^{u})\right] \,.
\end{split}
\end{equation}
We plug it in $A_1(t)$. 
As in the droplet calculation, the integral is convergent, but is composed of two diverging pieces, and we introduce the same convergence factor $e^{-\mu s}$ in both terms, with $0<\mu<1$. We take the limit only after subtracting them:
\begin{equation}\label{eq:flat_mureg}
\begin{split}
 & A_1(t)=\int_{\mathbb R}ds\,\left[\Theta(s)+e^s\mathcal Z_1(s+t/12,t)\right] \equiv \lim_{\mu\to0^+}\left[I_\mu^{\rm F}-\II_\mu^{\rm F}\right],\\
 & I_\mu^{\rm F}:=\int_{\mathbb R}ds\,e^{-\mu s}\Theta(s)=\frac1\mu,\qquad
 \II_\mu^{\rm F}:=-\int_{\mathbb R}ds\,e^{(1-\mu)s}\mathcal Z_1(s+t/12,t).
\end{split}
\end{equation}
The superscript distinguishes these integrals from their droplet counterparts. Without the regulator, the two terms are separately divergent: $-e^s\mathcal Z_1(s+t/12,t)\to1$ as $s\to+\infty$.
Inserting Eq.~\eqref{eq:Z1flat}, and exchanging the integrals at fixed $\mu$ gives
\begin{equation}\label{eq:flat_II_insert}
 \II_\mu^{\rm F}=\frac1{2(t/4)^{1/3}}\int_{\mathbb R}du\,[1-e^{-2e^u}]
 \int_{\mathbb R}ds\,e^{(1-\mu)s}
 {\rm Ai}\!\left(\frac{s+t/12+u}{(t/4)^{1/3}}\right).
\end{equation}
For the inner integral over $s$
, we substitute $s=(t/4)^{1/3} y-t/12-u$ and then use the Airy identity
\begin{equation}\label{eq:int1}
 \int_{\mathbb R}dy\,{\rm Ai}(y)e^{ay}=e^{a^3/3},\quad a>0,
\end{equation}
with $a=(1-\mu)(t/4)^{1/3}$, yielding
\begin{equation}\label{eq:flat_II_airy}
 \II_\mu^{\rm F}=\frac12
 \exp\!\left[\frac{t}{12}\big((1-\mu)^3-(1-\mu)\big)\right]
 \int_{\mathbb R}du\,e^{-(1-\mu)u}[1-e^{-2e^u}].
\end{equation}
The remaining integral is evaluated by $x=2e^u$, followed by integration by parts:
\begin{equation}\label{eq:flat_uint}
 \int_{\mathbb R}du\,e^{-(1-\mu)u}[1-e^{-2e^u}]
 =2^{1-\mu}\int_0^\infty dx\,x^{\mu-2}(1-e^{-x}) = \frac{2^{1-\mu}}{1-\mu}\int_0^\infty dx\,x^{\mu-1}e^{-x}
 =\frac{2^{1-\mu}\Gamma(\mu)}{1-\mu}.
\end{equation}
The boundary term vanishes at both endpoints precisely for $0<\mu<1$. This range also makes the double integral in Eq.~\eqref{eq:flat_II_insert} absolutely convergent, justifying the exchange of integrations.
Collecting the factors and expanding at fixed $t$, we obtain
\begin{equation}\label{eq:flat_IImu}
 \II_\mu^{\rm F} =\frac{2^{-\mu}\Gamma(\mu)}{1-\mu}
 \exp\!\left[\frac{t}{12}\big((1-\mu)^3-(1-\mu)\big)\right] =\frac1\mu+1-\gamma_{\rm E}-\ln2-\frac t6+O(\mu) \,.
\end{equation}
Here $\Gamma(\mu)=1/\mu-\gamma_{\rm E}+O(\mu)$, $2^{-\mu}/(1-\mu)=1+\mu(1-\ln2)+O(\mu^2)$, and $[(1-\mu)^3-(1-\mu)]t/12=-\mu t/6+O(\mu^2)$. Thus the poles cancel and
\begin{equation}\label{eq:flat_A1}
 A_1(t)=\frac t6+\ln2+\gamma_{\rm E}-1 \,.
\end{equation}
Adding back the first integral $1-\gamma_{\rm E}$ in Eq.~\eqref{eq:flat_def} gives the one-string result
\begin{equation}\label{eq:flat_one_string_result}
 S_{\rm F}(t)=1-\gamma_{\rm E}+A_1(t)+ A_2(t) + \dots   =\frac t6+\ln2+A_2(t) +\dots \,.
\end{equation}

\subsubsection{Subleading order: the two-string term}

In this paragraph, let us set $\mathcal Z_{\n}(s,t) \equiv \tilde{\mathcal Z}_{\n}(s/\lambda,\lambda)$ ($\equiv Z(\n,s/\lambda)$ in the notation of~\cite{le2012kpz}), with $\lambda=(t/4)^{1/3}$ for the $\n$-string term in the generating function $g_{\rm F}(s,t)$, which leads to a more compact notation. For instance, with this convention, Eq.~\eqref{eq:Z1flat} for the one-string term can be rewritten as
\begin{equation}
    \tilde{\mathcal Z}_1(s, \lambda) = - \frac12\int_{\mathbb R} du \ {\rm Ai} \left(s+u\right) \left[ 1-\exp(-2e^{\lambda u}) \right] \,,
\end{equation}
which matches with Eq.~\eqref{eq:Z1flat}. 
The two-string contribution is instead more involved. Define the two functions
\begin{equation} \label{eq:fdef}
\begin{split}
& f_k(z) = \sum_{m=1}^{\infty}
\prod_{q=1}^{m}
\frac{1}{4k^2+q^2}
(-z)^m = -z\,
\frac{{}_1F_2\!\left(1;2-2ik,\,2+2ik;-z\right)}
{4k^2+1} \,,\\ 
& F(z_1,z_2) =
\sinh(z_2-z_1)
+e^{-z_2} -e^{-z_1}
+\int_0^1 \mathrm{d}u\,
J_0\!\left(2\sqrt{z_1z_2u}\right)
\left[
z_1\sinh\!\left(z_1(1-u)\right)
-z_2\sinh\!\left(z_2(1-u)\right)
\right] \,.
\end{split}
\end{equation}
We then define the kernel 
\begin{equation} \label{eq:kernel}
\begin{split}
    K_{\lambda,11}(s;v_1,v_2) = \int_{\mathbb R^3} dy_1\ dy_2\ \frac{dk}{2\pi} \  & {\rm{Ai}}(y_1+v_1+s+4k^2) \ {\rm{Ai}}(y_2+v_2+s+4k^2) \times \\ 
    \times & \left [ \frac{\sin (2k(v_2-v_1))}{2k} f_{k/\lambda}(4e^{\lambda(y_1+y_2)}) + \frac 1 4 2\pi \delta(k) F(2e^{\lambda y_1},2e^{\lambda y_2})\right] \,,
\end{split}
\end{equation}
in terms of which we write the two-string  ``partition function'' (see Eq.~(153) of Ref.~\cite{le2012kpz} for comparison)
\begin{equation}\label{eq:Z2flat}
    \tilde{\mathcal Z}_2(s,\lambda) = -2 \int_0^\infty dv_1 \left. \partial_{v_1}  K_{\lambda, 11}(s;v_1,v_2) \right|_{v_1=v_2} \,.
\end{equation}
We can split the function $\tilde{\mathcal Z}_2(s,\lambda)$ into two contributions $\tilde{\mathcal Z}_2(s,\lambda) = \tilde{\mathcal Z}_2^I(s,\lambda) +\tilde{\mathcal Z}_2^{\II}(s,\lambda)$, with
\begin{equation} \small 
\begin{split}
    & \tilde{\mathcal Z}_2^I(s,\lambda) := 2 \int_{\mathbb R^3} dy_1 dy_2 \frac{dk}{2\pi} \int_0^\infty dv  {\rm{Ai}}(y_1+v+s+4k^2){\rm{Ai}}(y_2+v+s+4k^2) f_{k/\lambda} (4e^{\lambda(y_1+y_2)}) \\ 
    & \tilde{\mathcal Z}_2^{\II}(s,\lambda) := - \frac 1 4 \int_{\mathbb R^2} dy_1 dy_2 \int_0^\infty dv \Big[ {\rm{Ai}}'(y_1+v+s){\rm{Ai}}(y_2+v+s)  -{\rm{Ai}}(y_1+v+s){\rm{Ai}}'(y_2+v+s) \Big] F(2e^{\lambda y_1},2e^{\lambda y_2})\,,
\end{split} 
\end{equation}
where we used that $F(a,b)=-F(b,a)$. These expansions for the $2$-string term can be plugged into the integral $A_2(t)$ in Eq.~\eqref{eq:flat_def}, namely
\begin{equation} 
    A_2(t)= \frac12\int_{\mathbb R} ds\  e^s \mathcal{Z}_2(s+t/12,t) = \frac{e^{-t/12}}{2} \int_{\mathbb R} ds\  e^s \mathcal{Z}_2(s,t) = \frac{e^{-t/12}}{2} \left(\frac{t}{4} \right)^{1/3} \int_{\mathbb R} ds\ e^{s(t/4)^{1/3}} \tilde{\mathcal Z}_2\left(s, (\frac t 4)^{1/3}\right)\,.
\end{equation}
We rewrite $A_2 = J^{I}+J^{\II}$ ($t$-dependence implied), with 
\begin{equation}
    J^I :=\frac{e^{-t/12}}{2} \left(\frac{t}{4} \right)^{1/3} \int_{\mathbb R} ds\ e^{s(t/4)^{1/3}} \tilde{\mathcal Z}^I_2\left(s, (\frac t 4)^{1/3}\right)\,,\quad 
    J^{\II} :=\frac{e^{-t/12}}{2} \left(\frac{t}{4} \right)^{1/3} \int_{\mathbb R} ds\ e^{s(t/4)^{1/3}} \tilde{\mathcal Z}_2^{\II}\left(s, (\frac t 4)^{1/3}\right)\,.
\end{equation}

\textbf{Integral $J^I$.} --- Inserting the definition for $\tilde{\mathcal Z}_2^I$, the integral $J^I$ can be resolved in terms of the Airy integral identities already presented, which remove the $s$ integral. Integrating the remaining exponential over $v$, $J^I$ reduces to ($t$-dependence implied)
\begin{equation}
    J^I = \frac{1}{\sqrt{8}} \left( \frac{4}{t} \right)^{5/6} \frac{e^{-t/16}}{ \sqrt{4 \pi (\frac t 4)^{1/3}}} \int_{\mathbb R^3} dy_1 dy_2 \frac{dk}{2\pi}  e^{-\frac{y_1+y_2}{2}} e^{-k^2/2} e^{-\frac{1}{t}  (y_2-y_1)^2} f_{k/\sqrt{2t}}(4e^{y_1+y_2}) \,.
\end{equation}
Using the change of variable $u=(y_1+y_2)/2$ and the gap $y_2-y_1$, one has that the integral over the gap is Gaussian, and we obtain 
\begin{equation}
    J^I = \frac{e^{-t/16}}{\sqrt{2 t}} \int_{\mathbb R^2} du  \frac{dk}{2\pi}  e^{-u} e^{-k^2/2}  f_{k/\sqrt{2t}}(4e^{2u}) \,.
\end{equation}
From the definition of the function $f$ in Eq.~\eqref{eq:fdef} we compute 
\begin{equation}
    -\frac12\int_{0}^{\infty} \frac{dz}{z^2} f_{\lambda}(z^2) =  \frac12\frac{4 \pi \lambda}{1 + 16 \lambda^2} \coth 2 \pi \lambda =:  \mathsf g(\lambda) \,,
\end{equation}
and finally we insert $\mathsf g$ into the remaining integral over $k$ for $J^I$, which leads to the large-$t$ asymptotic expansion
\begin{equation}
    J^I(t) = - \frac{2 e^{-t/16}}{\sqrt{\pi t}}  \int_{\mathbb R} \frac{dk}{\sqrt{2\pi}} \ e^{-k^2/2}\ \mathsf g\left(\frac{k}{\sqrt{2 t}}\right) \simeq 
    - \frac{2 e^{-t/16}}{\sqrt{\pi t}}\Big[ 1 - \frac1t(8 - \frac{2}{3}\pi^2) + \frac{4}{15\ t^2} \left( 720 - 60 \pi^2 - \pi^4  \right) + O(t^{-3}) \Big] \,.
\end{equation}

\textbf{Integral $J^{\II}$.} --- Inserting the definition for $\tilde{\mathcal Z}_2^{\II}$, the derivatives of Airy functions can be removed by integration by parts. The $s$-integral is then performed by Airy identities, and the integral over $v$ is again a simple exponential. We are left with 
\begin{equation}
    J^{\II}  = -  \frac{e^{-t/16}}{8 \sqrt{4\pi (\frac t 4)^{1/3}}} \int_{\mathbb R^2} dy_1\ dy_2\ e^{-(\frac t 4 )^{1/3} \frac{y_1+y_2}{2}} \left[ (\partial_{y_1}-\partial_{y_2}) e^{-\frac 1 4 (\frac 4 t )^{1/3}(y_2-y_1)^2} \right] F(2e^{(\frac t 4)^{1/3}y_1},2e^{(\frac t 4)^{1/3}y_2}) \,.
\end{equation}
Scaling $y_{1/2} \mapsto (4/t)^{1/3}\, y_{1/2}$ and differentiating explicitly, we obtain
\begin{equation}
    J^{\II} = - \left( \frac{4}{t} \right)^{1/3} \frac{ e^{-t/16}}{2 t \sqrt{4\pi (\frac t 4)^{1/3}}} \int_{\mathbb R^2}\ dy_1 dy_2 (y_2-y_1) e^{- \frac{y_1+y_2}{2}}  e^{-\frac 1 t (y_2-y_1)^2}  F(2e^{y_1},2e^{y_2}) =
    -\frac{e^{-t/16}}{2\sqrt{\pi} t^{3/2}}  \mathsf J \,,
\end{equation}
where we introduced 
\begin{equation}
    \mathsf J(t) := \int_{\mathbb R^2} dy_1 dy_2\ (y_2-y_1)  e^{-\frac{y_1+y_2}{2}} e^{-\frac 1 t (y_2-y_1)^2} F(2e^{y_1},2e^{y_2}) \, .
\end{equation}
To compute $\mathsf J$, we exploit the fact that the Laplace transform of $F(z_1,z_2)$ can be written as
\begin{equation}
    \mathcal{L}[F(z_1,z_2)] (s_1,s_2) = \frac{1}{s_1 s_2} \frac{1}{(1+s_2)(1+s_1)} \frac{s_2-s_1}{1+s_1 s_2} \,,
\end{equation}
so that one has 
\begin{equation}
    \mathsf J(t) = \frac{1}{(2 \pi i)^2} \int_{\mathbb R^2} dy_1\ dy_2\ (y_2-y_1) e^{-\frac 1 t (y_2-y_1)^2} e^{-\frac{y_1+y_2}{2}}  \int_\Gamma \frac{ds_1\ ds_2}{s_1 s_2}\ \frac{e^{2 s_1 e^{y_1} + 2 s_2 e^{y_2}}}{(1+s_2)(1+s_1)} \frac{s_2-s_1}{1+s_1 s_2} \,,
\end{equation}
where $\Gamma$ is an appropriate complex contour parallel to the imaginary axis. Now, we first rescale $s_i\mapsto e^{-y_i}s_i$, and then shift $y_i=Y_i+\ln s_i$, using the corresponding contour deformations. In particular,
\begin{equation}
 e^{-y_i}s_i=e^{-Y_i},\qquad
 e^{-(y_1+y_2)/2}=\frac{e^{-(Y_1+Y_2)/2}}{(s_1s_2)^{1/2}},\qquad
 y_2-y_1=Y_2-Y_1+\ln \frac{s_2}{s_1}.
\end{equation}
and the integral therefore reads
\begin{equation}
\begin{split}
 \mathsf J(t)=&\frac{1}{(2\pi i)^2}\int_\Gamma
 \frac{ds_1ds_2}{(s_1s_2)^{3/2}}e^{2(s_1+s_2)}
 \int_{\mathbb R^2}dY_1\,dY_2\,
 \left[Y_2-Y_1+\ln \frac{s_2}{s_1}\right]\\
 &\times e^{-\tfrac1t[Y_2-Y_1+\ln\tfrac{s_2}{s_1}]^2}
 e^{-\tfrac12(Y_1+Y_2)}
 \frac{e^{-Y_2}-e^{-Y_1}}
 {(1+e^{-Y_2})(1+e^{-Y_1})(1+e^{-Y_1-Y_2})} \,.
\end{split}
\end{equation}
Now set $Y_1=u-v/2$, $Y_2=u+v/2$, with unit Jacobian. Interchanging the dummy contour variables $s_1,s_2$ lets us write the shifted difference as $v-\ell$, with $\ell:=\ln s_2-\ln s_1$, giving
\begin{equation}
\label{eq:Jt}
\mathsf J(t)=\frac{1}{(2\pi i)^2}\int_\Gamma
 \frac{ds_1ds_2}{(s_1s_2)^{3/2}}e^{2(s_1+s_2)}
 \int_{\mathbb R}dv\,(v-\ell)e^{-(v-\ell)^2/t}H(v) \,,
\end{equation}
with the function $H$ defined by the integral over $u$:
\begin{equation}
    H(v):=\int_{\mathbb R}du\,
 \frac{e^{-v/2}-e^{v/2}}
 {(1+e^{-u-v/2})(1+e^{-u+v/2})(1+e^{2u})} \,.
\end{equation}
The substitution $x=e^u$ reduces the remaining $u$-integral to an elementary rational integral:
\begin{equation}    H(v)=-2\sinh(v/2)\int_0^\infty
 \frac{x\,dx}{(x+e^{-v/2})(x+e^{v/2})(1+x^2)}
 =\frac12\left[\frac{v}{\cosh(v/2)}-\pi\tanh(v/2)\right] \,.
\end{equation}
Finally, we use $(v-\ell)e^{-(v-\ell)^2/t}=-(t/2)\partial_v e^{-(v-\ell)^2/t}$ and integrate by parts in $v$ in Eq.~\eqref{eq:Jt}. The boundary term vanishes, and using
\begin{equation}
     H'(v)=-\frac{1}{2\cosh^2(v/2)}
 \left[\frac\pi2-\cosh(v/2)+\frac v2\sinh(v/2)\right] \,,
\end{equation}
we arrive at the expression
\begin{equation}
    \mathsf J(t) =-\frac{t}{4(2 \pi i)^2} \int_\Gamma \frac{ds_1 ds_2}{(s_1 s_2)^{3/2}} e^{2 (s_1  + s_2)} \int_{\mathbb R} \frac{dv}{\cosh(v/2)^2} \  e^{-\left(v - \ln(s_2/s_1) \right)^2/t}  \left( \frac{\pi}{2} - \cosh \frac{v}{2} + \frac{v}{2} \sinh \frac{v}{2}  \right) \,.
\end{equation}
By expanding in powers of $1/t$, one can perform the integrals over $s_{1,2}$ and $v$, leading to
\begin{equation}
    \mathsf J(t) = \frac{t}{8 \pi} z(\alpha)^2 \left[ 1 - \left( \frac{7}{3} \pi^2 + 2 \partial^2_\alpha \ln z \right) t^{-1} + \left(\frac{307 \pi^4}{30} + 14 \pi^2 \partial_\alpha^2 \ln z + \partial_\alpha^4 \ln z + 6 (\partial_\alpha^2 \ln z)^2 \right) t^{-2} +  O(t^{-3}) \right]_{\alpha = 3/2} \,,
\end{equation}
where
\begin{equation}
    z(\alpha) = \int_\Gamma \frac{ds}{s^\alpha} e^{2s} = \frac{2^\alpha \pi}{\Gamma(\alpha)} i \,,
\end{equation}
so that 
\begin{equation}
    \mathsf J(t) = - 4t \left[ 1- \left( 8 + \frac{4}{3} \pi^2 \right) t^{-1} + \left( 192 +32 \pi^2 + \frac{56}{15} \pi^4 \right) t^{-2} + O(t^{-3}) \right] \,.
\end{equation}
We thus conclude 
\begin{equation}
    J^{\II}(t) =  \frac{2}{\sqrt{\pi t}} e^{-t/16} \left[ 1- \left( 8 + \frac{4}{3} \pi^2 \right) t^{-1} + \left( 192 +32 \pi^2 + \frac{56}{15} \pi^4 \right) t^{-2} + O(t^{-3}) \right] \,.
\end{equation}

\textbf{Total contribution.} --- Putting together $J^I$ and $J^{\II}$ into $A_2(t)$, the leading terms cancel. The contribution from the ``flat'' partition function to the entropy is given by the sum of integrals $A_1(t)$ and $A_2(t)$ together with the constant term, as in \eqref{eq:flat_def}, namely
\begin{equation} \label{eq:SFtotal}
    S_{\rm F}(t) = \frac{t}{6} + \ln 2 - 4 e^{-t/16} \left( \frac{\pi}{t} \right)^{3/2}  \left( 1 -  \frac{2}{t} (12 + \pi^2) + O(t^{-2}) \right)  + \ldots \,.
\end{equation}

\subsection{Total: approach to saturation}

Collecting Eqs.~(\ref{eq:SDwithI},~\ref{eq:SFtotal}), the extensive parts $t/6$ cancel and
\begin{equation}
    \s[t] = S_{\rm F}(t)-S_{\rm D}(t) = \underbrace{1+\gamma_{\rm E}+\ln 2}_{S_\infty} - 4 e^{-t/16} \left( \frac{\pi}{t} \right)^{3/2}  \left( 1 -  \frac{2(12 + \pi^2)}{t} + \frac{16\pi^{3/2}}{t^{3/2}}+O(t^{-2}) \right)  + \ldots \,,
\end{equation}
which is Eq.~\eqref{eq:approach} of the main text.

\section{Power-law $c(t)$ with $\alpha>1/2$: perturbation theory around free diffusion} \label{sec:alphamax}

This Section details the computation sketched in the End Matter, and extends it to $\alpha\geq3/2$.
Let us consider the SHE with time-dependent noise variance Eq.~\eqref{eq:SHE} of the main text, with $c(t)$ a function such that: (i) $c(0)=1$ at initial time, which can always be set by a standard change of variables $(x,t)$; (ii) at long times $t\geq t_0$, it is a power-law $c(t) = c_0 (t/t_0)^{-\alpha}$. We want to study the moments
$\average{\mathbf x|\Psi(t)} := \overline{\prod_{i=1}^n  z(x_i,t)}$ via the imaginary-time Schr\"odinger representation
\begin{equation}
    \langle \mathbf x | \Psi(t) \rangle = 
    \langle \mathbf x | \mathcal T e^{-\int_0^t dt' \hat H_{\rm LL}(t
    ')} | 0\rangle \,,\quad 
    \hat H_{\rm LL}(t) = \hat H_0 - 2c(t) \sum_{1\leq i<j\leq n} \delta(\hat x_i-\hat x_j) \,,\quad 
    \hat H_0 = -\sum_{i=1}^n \nabla_i^2\,,
\end{equation}
with the initial condition $\langle \mathbf x|0\rangle = \prod_{i=1}^n \delta(x_i)$. 
We assume that the behavior of $c(t)$ is unknown up to $t=t_0$, and we encode it in a state $\ket{\Psi(t_0)}$ such that 
\begin{equation}
    | \Psi(t) \rangle = 
    \mathcal T e^{-\int_{t_0}^t dt' \hat H_{\rm LL}(t
    ')} | \Psi(t_0) \rangle \,.
\end{equation}
It is convenient to move to the interaction picture
\begin{equation}
    \ket{\Psi_I(t)} = e^{\hat H_0 (t-t_0)} \ket{\Psi(t)} \,\Rightarrow\, 
    \ket{\Psi_I(t)} = \mathcal T e^{-\int_{t_0}^t dt' \hat V_I(t') } \ket{\Psi(t_0)} \,,\quad 
    \hat V_I(t) = -2c(t) \ e^{\hat H_0(t-t_0)} \sum_{i<j} \delta(x_i-x_j) e^{-\hat H_0(t-t_0)} \,,
\end{equation}
so that we then expand the time-evolution in perturbation theory, with correction terms in powers of $c_0$
\begin{equation}
\begin{split}
    & \ket{\Psi_I(t)} = \ket{\Psi^{(0)}_I(t)} + c_0 \ket{\Psi^{(1)}_I(t)}  + O(c_0^2) \,, \\
    & \mathcal T e^{-\int_{t_0}^t dt' \hat V_I(t') } = \mathbb I + 2c_0 t_0^\alpha \int_{t_0}^t dt' \ (t')^{-\alpha} \ e^{\hat H_0(t'-t_0)} \sum_{i<j} \delta(x_i-x_j) e^{-\hat H_0(t'-t_0)} + O(c_0^2) \,.
\end{split}
\end{equation}

\textbf{Zeroth-order terms.} --- The zeroth order reproduces the free-diffusion problem. Introducing the free propagator $G_n(\xvec;t):=e^{-\sum_{i} x_i^2/(4t)}/[4\pi t]^{n/2}$, 
by conservation of probability one has
\begin{equation}
\begin{split}
    & \int d^n x\, \langle \mathbf x\ket{\Psi^{(0)}(t)} = \int d^nx\ d^n y\ G_n(\xvec-\yvec ;t-t_0)\ \langle \yvec \ket{\Psi(t_0)}  = 1 \,,\\ 
    & \int dx\ \langle x\hat u\ket{\Psi^{(0)}(t)} = \int dx\ d^n y\ G_n(\xvec-\yvec ;t-t_0)\ \langle \yvec \ket{\Psi(t_0)}  = \frac{\int d^n y \ \langle \yvec | \Psi(t_0)\rangle \exp\left[-\frac{\sum_{i<j}(y_i-y_j)^2}{4n (t-t_0)}\right]}{\sqrt n [4\pi(t-t_0)]^{(n-1)/2}} \,.
\end{split}
\end{equation}
While the first integral in $d^n x$ is exact, the second integral on the diagonal $\xvec = x\hat u$, with $\hat u = (1,\dots,1)$, requires a full knowledge of the state $\ket{\Psi(t_0)}$. However, at large times $t\gg t_0$, for a localized initial state with finite relative-coordinate moments, we can expand the exponential kernel as 
\begin{equation}
    \int d^n y \ \langle \yvec | \Psi(t_0)\rangle \exp\left[-\frac{\sum_{i<j}(y_i-y_j)^2}{4 n (t-t_0)}\right] = 1 - \frac{1}{4 n (t-t_0)}\sum_{i<j}  \int d^n y \ \langle \yvec | \Psi(t_0)\rangle (y_i-y_j)^2 + O(t^{-2}) \,,
\end{equation}
and only the first term survives at large times. Therefore, while the leading $\frac12\ln t$ term \emph{and} the $O(1)$ constant are captured exactly, the coefficient for corrections $O(t^{-1})$ is determined by the early-time, non-perturbative evolution. Finally, this gives the zeroth-order result
\begin{equation}
    S^{(0)}(t) = \frac12 \left( 1+\ln 4\pi t \right) +O(t^{-1}) \,.
\end{equation}

\textbf{First-order terms.} --- Conversely, the first-order correction $\propto c_0$ encodes two-body scatterings 
\begin{equation}
    \langle \mathbf x\ket{\Psi^{(1)}(t)} = 
    2 t_0^\alpha \sum_{1\leq i<j\leq n} \int_{t_0}^t dt' (t')^{-\alpha} \bra{\xvec}e^{-\hat H_0(t-t')} \delta(x_i-x_j) e^{-\hat H_0(t'-t_0)} \ket{\Psi(t_0)} \,.
\end{equation}
We neglect the early-time broadening of the propagator as stated in the main text, and approximate $\langle \xvec \ket{\Psi(t_0)} \approx \prod_{i=1}^n \delta(x_i)$. 
Inserting completeness relations in the position basis we then proceed as 
\begin{equation}
\begin{split}
    \langle \mathbf x \ket{\Psi^{(1)}(t)} & =  2 t_0^\alpha\sum_{i<j} \int_{t_0}^t dt' (t')^{-\alpha} \int d^ny \ \bra{\xvec}e^{-\hat H_0(t-t')} \ket{\yvec} \delta(y_i-y_j) \bra{\yvec}e^{-\hat H_0(t'-t_0)} \ket{\Psi(t_0)} \\ 
    &\approx n(n-1) \ t_0^\alpha  \int_{t_0}^t dt' (t')^{-\alpha} \int d^ny \ G_n(\xvec-\yvec;t-t') \delta(y_1-y_2) G_n(\yvec;t'-t_0) = \\ 
    &= n(n-1) \ t_0^\alpha\ G_{n-2}(\xvec,t-t_0)  \int_{t_0}^t dt' (t')^{-\alpha} \int dy \ G_1(x_1-y,t-t')G_1(x_2-y,t-t')G_2(y,t'-t_0) \,.
\end{split}
\end{equation}
From here we can integrate over the Gaussian terms, and the flat and droplet terms yield, respectively:
\begin{equation}
\small 
\begin{split} 
    \int d^n x \ \langle \xvec| \Psi^{(1)}\rangle & \approx n(n-1)\  t_0^\alpha\ \int d^nx\  G_{n-2}(\xvec,t-t_0)  \int_{t_0}^t dt' t'^{-\alpha} \int dy \ G_1(x_1-y,t-t')G_1(x_2-y,t-t')G_2(y,t'-t_0) = \\ 
    & = n(n-1)\ \frac{t_0^\alpha}{\sqrt{8\pi}} \int_{t_0}^t dt' \frac{t'^{-\alpha}}{\sqrt{t'-t_0}} \,,
\end{split}
\end{equation}
\begin{equation}
\begin{split} 
    \int dx \ \langle x\hat u| \Psi^{(1)}(t)\rangle & \approx n(n-1)\ t_0^\alpha \int dx\  G_{n-2}(x \hat u,t-t_0)  \int_{t_0}^t dt' t'^{-\alpha} \int dy \ G_2(x-y,t-t') G_2(y,t'-t_0) =\\
    & = \frac{\sqrt n(n-1)}{[4\pi(t-t_0)]^{(n-1)/2}}\ \frac{t_0^\alpha}{\sqrt{8\pi}}  \int_{t_0}^t dt' \ \frac{(t')^{-\alpha}}{\sqrt{t'-t_0}}\sqrt{\frac{t-t_0}{t-t'}} \,.
\end{split}
\end{equation}
Importantly, both integrals in $t'$ converge at the edge $t'\to t_0$, while, for $\alpha\leq 1/2$, they do not converge as time diverges $t\to\infty$. This signals the breakdown of perturbation theory for this choice of $\alpha\leq 1/2$.
Conversely, for $\alpha>1/2$, both integrals converge at large $t$ to the \emph{same} constant $C_\infty=\frac{t_0^\alpha}{\sqrt{8\pi}}\int_{t_0}^\infty dt'\,t'^{-\alpha}(t'-t_0)^{-1/2}$, dominated by the $t'\approx t_0$ region. There $c(t')=O(1)$, so that the individual constants are \emph{not} controlled by late-time perturbation theory; the entropy, however, only involves their difference, which is dominated by $t'\sim t$ and is therefore perturbatively safe. Defining
\begin{equation}
    C_{\rm F}(t) = \frac{t_0^\alpha}{\sqrt{8\pi}}\int_{t_0}^{t} \frac{dt'\ t'^{-\alpha}}{\sqrt{t'-t_0}}\,,\quad
    C_{\rm D}(t) = \frac{t_0^\alpha}{\sqrt{8\pi}}\int_{t_0}^{t} \frac{dt'\ t'^{-\alpha}}{\sqrt{t'-t_0}}\ \sqrt{\frac{t-t_0}{t-t'}}\,,
\end{equation}
and using $\partial_n[n(n-1)]_{n=1}=1$, we find $S^{(1)} (t)\stackrel{t \gg t_0}{\approx}C_{\rm F}(t)-C_{\rm D}(t)$ yielding Eq.~\eqref{eq:firstterm} in the main text. The term in square brackets therein behaves as $\frac{t'-t_0}{2(t-t_0)}+\cdots$ for $t'-t_0\ll t-t_0$. 
We now evaluate the integral in $t'$. While the integral can be performed exactly in $t'$, it is convenient to set $t'=t u$, and in the difference the bracket cures the $u\to0$ region so that one may set $t_0/t\to0$. The integral scales as $t^{1/2-\alpha}$ for $\tfrac12<\alpha<\tfrac32$, which is dominant with respect to the $O(t^{-1})$ correction coming from the zeroth-order terms. Therefore:
\begin{equation}
    S^{(1)} (t) \approx -\frac{t_0^\alpha}{\sqrt{8\pi}}\ I(\alpha)\ t^{\frac12-\alpha}\,,\quad
    I(\alpha) = \int_0^1 du\ u^{-\alpha-\frac12}\Big[(1-u)^{-\frac12}-1\Big] = \sqrt{\pi} \ \frac{\Gamma(\tfrac12-\alpha)}{\Gamma(1-\alpha)}+\frac{1}{\alpha-1/2} > 0\,,
\end{equation}
which is finite, with a smooth marginal limit $\lim_{\alpha\to 1/2^+}I(\alpha)=2\ln2$ (the $\Gamma$-poles cancel at $\alpha\to\tfrac12^+$; substitute $w=\sqrt{1-u}$ to check the value), and
\begin{equation}
    \s[t] = S^{(0)}(t) + c_0 S^{(1)} (t) + O(c_0^2) =  \frac12[\ln 4\pi t +1] - \frac{c_0 t_0^\alpha}{\sqrt{8\pi}}\ I(\alpha)\ t^{\frac12-\alpha} + O(c_0^2) + O(t^{-1})\,,\quad \frac 12<\alpha<\frac 3 2 \,, 
\end{equation}
with no $\alpha$-dependent $O(1)$ constant: the entropy approaches free diffusion from below, with a reduction that dies out faster for larger $\alpha$. 

For $\alpha=3/2$, the asymptotics of the integral features $\log$ corrections, and it is possible to compute the coefficient as 
\begin{equation}
    \int_{t_0}^{t} \frac{dt'\ t'^{-\alpha}}{\sqrt{t'-t_0}}\left[\sqrt{\frac{t-t_0}{t-t'}}-1\right] = \frac{\ln (t/t_0)}{2t}+\frac{4 \ln2 - 1}{2t} +O(t^{-2}) \,,
\end{equation}
hence 
\begin{equation}
    \s[t] = \frac12[\ln 4\pi t +1]  - \frac{c_0 t_0^\alpha}{\sqrt{8\pi}} \frac{\ln (t/t_0)}{2t} + O(c_0^2) +O(t^{-1}) \,,
\end{equation}
where the $O(t^{-1})$ terms are dropped as they have to be compared with the corrections coming from the exact shape of the initial state $\ket{\Psi(t_0)}$. 

Finally, for $\alpha>3/2$ the corrections coming from 
\begin{equation}
    C_{\rm D}(t)-C_{\rm F}(t) = \frac{t_0^{3/2}}{8\sqrt 2}\frac{\Gamma(\alpha-\frac32)}{\Gamma(\alpha)}\frac1t + O(t^{-2})\,,\quad \alpha>\frac3 2 \,,
\end{equation}
are of the same order as those coming from the zeroth-order propagator, therefore 
\begin{equation}
    \s[t] = \frac12[\ln 4\pi t +1]  +O(t^{-1})\,,\quad \alpha>\frac3 2 \,.
\end{equation}

\section{Power-law $c(t)$ with $0<\alpha<1/2$: the Lieb-Liniger propagator in a harmonic potential} \label{sec:LLperturb}

\subsection{Center-of-mass kernel}

We first give the explicit solution of the center-of-mass problem Eq.~\eqref{eq:COMeq} of the End Matter, which is not needed for the entropy --- only the conservation law~\eqref{eq:COM} is --- but which shows that the center of mass is exactly solvable for an arbitrary profile $A(\tau)$. Since the potential is quadratic, the solution is Gaussian and can be written in closed form in terms of two auxiliary functions. Let $u(\tau)$ and $v(\tau)$ solve
\be
f''(\tau)=2A(\tau) f(\tau) = \frac{c''(\tau)}{c(\tau)} f(\tau)\,,
\label{eq:uvode}
\ee
with initial conditions $u(0)=0,\;u'(0)=1$, and $v(0)=1,\;v'(0)=0$. Then the solution of \eqref{eq:COMeq} with initial condition $K_n(R,0;R_0)=\delta(R-R_0)$ is
\be
K_n(R,\tau;R_0)
=\sqrt{\frac{n}{4\pi u(\tau)}}
\exp\!\left[-\frac{n}{4u(\tau)}\,Q(R,\tau ;R_0)\right] \,,\quad
Q(R,\tau;R_0):= u'(\tau)R^2+v(\tau)R_0^2-2RR_0 \,.
\label{eq:COMkernel}
\ee
Note that the Wronskian $f'(\tau) c(\tau) - f(\tau) c'(\tau)$ is constant for any solution $f$; in particular $u'c-uc'=1$. Using this in \eqref{eq:COMkernel} one recovers Eq.~\eqref{eq:COM},
\begin{equation}
    \int \frac{d R}{\sqrt{c(\tau)}} \ e^{n\frac{c'(\tau)}{4c(\tau)}R^2} K_n(R,\tau;0)
     = \int \sqrt{\frac{n}{4\pi c u}}\, dR\ e^{-\frac{n R^2}{4 u c}(u' c - u c')} = 1\,,
\label{eq:COMexplicit}
\end{equation}
consistently with the direct proof given in the End Matter, which does not use the explicit form of $K_n$.

\subsection{Corrections to the relative propagator}

This Subsection justifies Eq.~\eqref{eq:GSproj} of the main text. The relative propagator $k_n(\tyvec,\tau)=\langle\tyvec\ket{\check\Psi(\tau)}$ singled out by the factorization Eq.~\eqref{eq:COMREL} is the only object entering the moments Eq.~\eqref{eq:momentsmapped}, and hence the entropy Eq.~\eqref{eq:Sdiffder}; the center-of-mass dynamics is resolved exactly and plays no role below. Using $\sum_i y_i^2=nR^2+\f(\tyvec)$ in Eq.~\eqref{eq:LLharmonic}, $\ket{\check\Psi(\tau)}$ evolves according to
\begin{equation} \label{eq:relharmonic}
    \partial_\tau \ket{\check\Psi(\tau)} = -\Big[\hat H_{\rm LL}(1) + \frac{A(\tau)}{2}\,\f(\tyvec)\Big]\ket{\check\Psi(\tau)} \,,
\end{equation}
where $\hat H_{\rm LL}(1)$ now acts on the relative coordinates $\tyvec\in\mathbb R^{n-1}$, with initial condition $\langle\tyvec\ket{\check\Psi(\tau=0)}=\prod_{i=1}^{n-1}\delta(\tilde y_i)$. The roles are exchanged with respect to Sec.~\ref{sec:alphamax}: the unperturbed problem is the \emph{interacting} LL Hamiltonian, and the perturbation is the trap, whose amplitude Eq.~\eqref{eq:potentialdef} we write as $A(\tau)=a\,(\tau-\tau_0)^{-2}$ with $a:=\alpha(1-\alpha)/[2(1-2\alpha)^2]$. The trap is not small at $\tau(t_0)$, where $\tau(t_0)-\tau_0=c_0^2t_0/(1-2\alpha)$ and $A=\alpha(1-\alpha)/(2c_0^4t_0^2)$ is in fact large for small $c_0$. We therefore evolve exactly up to a reference time $\tau_*>\tau(t_0)$ such that $A(\tau_*)\ll1$, and expand around $\ket{\check\Psi(\tau_*)}$; as we will see, the arbitrariness of $\tau_*$ only affects the nonuniversal amplitude $\kappa_n$. To first order in the trap, the Dyson expansion gives
\begin{equation} \label{eq:Dyson1}
    \ket{\check\Psi(\tau)} = \underbrace{e^{-\hat H_{\rm LL}(1)(\tau-\tau_*)}\ket{\check\Psi(\tau_*)}}_{\ket{\check\Psi^{(0)}(\tau)}}
    \ \underbrace{-\int_{\tau_*}^{\tau}\! d\tau'\,\frac{A(\tau')}{2}\; e^{-\hat H_{\rm LL}(1)(\tau-\tau')}\,\f\; e^{-\hat H_{\rm LL}(1)(\tau'-\tau_*)}\ket{\check\Psi(\tau_*)}}_{\ket{\check\Psi^{(1)}(\tau)}} + O(A^2)\,,
\end{equation}
the first-order term being a single scattering off the trap at time $\tau'$, with unperturbed LL propagation before and after it. The expansion is controlled by $A(\tau_*)\ll1$, measured in the units set by the gap of $\hat H_{\rm LL}(1)$, which, as we show, is finite at finite $n$.

\textbf{Spectrum and zeroth order.} --- As already discussed at the beginning of Subsec.~\ref{subsec:flatpolym}, the spectrum of $\hat H_{\rm LL}(1)$ is exactly known: eigenstates are labelled by a
partition $\{n_j\}_{j=1}^{\n}$ of $n$ into ``strings'' with real centers $k_j$, and
\begin{equation} \label{eq:stringspectrum}
    E = \sum_{j=1}^{\n} n_j k_j^2 - \frac{1}{12}\sum_{j=1}^{\n} n_j(n_j^2-1) \,,
    \qquad P = \sum_{j=1}^{\n} n_j k_j \,,\qquad n = \sum_{j=1}^{\n} n_j \,,
\end{equation}
where $P$ is the total momentum, fixed to $P=0$ in the relative sector. The ground state is the bound state of all $n$ particles, $\n=1$, $n_1=n$, $k_1=0$, with $E_n^{\rm GS}=-n(n^2-1)/12$; we take $\ket{\Psi_n^{\rm GS}}$ normalized in $L^2(\mathbb R^{n-1})$, with $|| \Psi|| := \sqrt{\braket{\Psi}{\Psi}}$. The excited states are obtained by breaking this bound state, and the lowest threshold is the $(n-1,1)$ break-up, $n_1=n-1$ and $n_2=1$ with $P = (n-1)k_1+k_2=0$: a continuum starting at $k_1=k_2=0$, i.e.\ at a distance
\begin{equation} \label{eq:gap}
    \Delta_{\rm gap}(n) := \frac{n(n^2-1)-(n-1)[(n-1)^2-1]}{12} = \frac{n(n-1)}{4} > 0
    \,, \qquad n\geq 2 \,,
\end{equation}
above the ground state. Introducing the projectors $\Pi:=\ket{\Psi_n^{\rm GS}}\bra{\Psi_n^{\rm GS}}$ and $\Pi^\perp:=\mathbb I-\Pi$, which commute with $\hat H_{\rm LL}(1)$, the gap implies $\|\Pi^\perp e^{-\hat H_{\rm LL}(1)s}\ket{v}\|\leq e^{-(E_n^{\rm GS}+\Delta_{\rm gap})s}||v||$ for every $s\geq0$, so that the zeroth-order term projects onto the ground state,
\begin{equation} \label{eq:zeroth}
\begin{split}
    &\langle \tyvec \ket{\check\Psi^{(0)}(\tau)} = \kappa_n^{(0)} e^{-E_n^{\rm GS}\tau}\left[ \psi_n(\tyvec) + O\big(e^{-\Delta_{\rm gap}(\tau-\tau_*)}\big)\right] \,,\\
    &\psi_n(\tyvec) := \langle \tyvec\ket{\Psi_n^{\rm GS}}\average{\Psi_n^{\rm GS}|\check\Psi(\tau=0)} \,,\qquad
    \kappa_n^{(0)} := e^{E_n^{\rm GS} \tau_*}\frac{\c_0}{\average{\Psi_n^{\rm GS}|\check\Psi(\tau=0)}} \,,
\end{split}
\end{equation}
with $\c_0:=\average{\Psi_n^{\rm GS}\big|\check\Psi(\tau_*)}$, which is Eq.~\eqref{eq:GSproj} at the lowest order in large $\tau$. The first-order term $\ket{\check\Psi^{(1)}}$ can alter it in two ways, which we examine in turn: its component along $\ket{\Psi_n^{\rm GS}}$ corrects the amplitude $\kappa_n$, its component orthogonal to it deforms the shape $\psi_n$.

\textbf{Projection on the ground state.} --- Projecting Eq.~\eqref{eq:Dyson1} on $\bra{\Psi_n^{\rm GS}}$, the propagation after the scattering reduces to $e^{-E_n^{\rm GS}(\tau-\tau')}$, and
\begin{equation} \label{eq:Pfirst}
    \average{\Psi_n^{\rm GS}\Big|\check\Psi^{(1)}(\tau)} = -e^{-E_n^{\rm GS}(\tau-\tau_*)} \int_{\tau_*}^\tau d\tau'\, \frac{A(\tau')}{2}\ e^{E_n^{\rm GS}(\tau'-\tau_*)}\bra{\Psi_n^{\rm GS}}\f\, e^{-\hat H_{\rm LL}(1)(\tau'-\tau_*)}\ket{\check\Psi(\tau_*)} \,.
\end{equation}
We split the incoming state as $e^{-\hat H_{\rm LL}(1)(\tau'-\tau_*)}\ket{\check\Psi(\tau_*)} = \c_0 e^{-E_n^{\rm GS}(\tau'-\tau_*)}\ket{\Psi_n^{\rm GS}} + \Pi^\perp e^{-\hat H_{\rm LL}(1)(\tau'-\tau_*)}\ket{\check\Psi(\tau_*)}$. The ground-state part contributes $\f_{00}\,\c_0\int_{\tau_*}^\tau d\tau' A(\tau')/2=\f_{00}\,\c_0\,\tfrac a2\big[(\tau_*-\tau_0)^{-1}-(\tau-\tau_0)^{-1}\big]$, with $\f_{00}:=\bra{\Psi_n^{\rm GS}}\f\ket{\Psi_n^{\rm GS}}$. The excited part is damped by $e^{-\Delta_{\rm gap}(\tau'-\tau_*)}$, with a prefactor $\|\f \ket{\Psi_n^{\rm GS}}\|\,\|\ket{\check\Psi(\tau_*)}\|$ which is finite because $\Psi_n^{\rm GS}$ decays exponentially in every relative coordinate while $\f$ is only quadratic; its integral therefore converges, up to $O(e^{-\Delta_{\rm gap}(\tau-\tau_*)})$, to the $\tau$-independent constant
\begin{equation} \label{eq:Aconst}
    \mathcal A := \int_{\tau_*}^\infty d\tau'\, \frac{A(\tau')}{2}\ e^{E_n^{\rm GS}(\tau'-\tau_*)}\bra{\Psi_n^{\rm GS}}\f\,\Pi^\perp e^{-\hat H_{\rm LL}(1)(\tau'-\tau_*)}\ket{\check\Psi(\tau_*)} \,.
\end{equation}
Using $\langle\tyvec\ket{\Psi_n^{\rm GS}}\,\c_0\,e^{-E_n^{\rm GS}(\tau-\tau_*)}=\kappa_n^{(0)}e^{-E_n^{\rm GS}\tau}\psi_n(\tyvec)$, the component of $\ket{\check\Psi(\tau)}$ along the ground state is thus $\kappa_n(\tau)\,e^{-E_n^{\rm GS}\tau}\psi_n(\tyvec)$ with
\begin{equation} \label{eq:kappat}
    \kappa_n(\tau) = \kappa_n^{(0)}\left[1 - \frac{a\,\f_{00}}{2(\tau_*-\tau_0)} - \frac{\mathcal A}{\c_0}\right] + \kappa_n^{(0)}\,\frac{a\,\f_{00}}{2(\tau-\tau_0)} + O(A^2)\,.
\end{equation}
The constant bracket is a renormalization of the nonuniversal amplitude $\kappa_n^{(0)}$, and it is where the dependence on  $\tau_*$ survives. The $O(\tau^{-1})$ term comes from the ground-state part of the incoming state, for which the scattering pays no exponential penalty and the trap integrates over the entire interval.

\textbf{Projection on the orthogonal subspace.} --- Projecting Eq.~\eqref{eq:Dyson1} on $\Pi^\perp$ instead, the propagation \emph{after} the scattering lies in the excited subspace and is damped by the gap, whereas the propagation before it is not, since there the state is still dominated by the ground state. Hence
\begin{equation} \label{eq:Qbound}
    \left\|\Pi^\perp\ket{\check\Psi^{(1)}(\tau)}\right\| \leq \frac{M}{2}\, e^{-E_n^{\rm GS}(\tau-\tau_*)} \int_{\tau_*}^\tau d\tau'\, A(\tau')\, e^{-\Delta_{\rm gap}(\tau-\tau')} \,,\qquad
    M := \sup_{s\geq0}\ e^{E_n^{\rm GS}s}\,\left\|\f\, e^{-\hat H_{\rm LL}(1)s} \ket{\check\Psi(\tau_*)}\right\| \,.
\end{equation}
The quantity $M$ is finite as it involves the supremum over a continuous function of $s$
which equals
$\|\f\ket{\check\Psi(\tau_*)}\|$ at $s=0$ (finite since $\ket{\check\Psi(\tau_*)}$ has Gaussian tails for $\tau_*>0$), and tends to $|\c_0|\,\|\f\ket{\Psi_n^{\rm GS}}\|$ as $s\to\infty$.

Physically, the attractive LL evolution binds the replicas instead of spreading them, so the second moment of the relative wavefunction stays bounded. The $\tau'$-integral in Eq.~\eqref{eq:Qbound} is dominated by the last interval of duration $\sim\Delta_{\rm gap}^{-1}$ before $\tau$, where $A(\tau')\simeq A(\tau)$: it equals $A(\tau)\,\Delta_{\rm gap}^{-1}\big[1+O((\Delta_{\rm gap}\tau)^{-1})\big]$ up to terms $O(e^{-\Delta_{\rm gap}(\tau-\tau_*)/2})$ from early scatterings, which meet a stronger trap but must then survive a long excited propagation. Therefore
\begin{equation} \label{eq:Qresult}
    \left\|\Pi^\perp\ket{\check\Psi^{(1)}(\tau)}\right\| \leq \frac{M}{2\Delta_{\rm gap}}\, A(\tau)\, e^{-E_n^{\rm GS}(\tau-\tau_*)}\left[1+O\big((\Delta_{\rm gap}\tau)^{-1}\big)\right] = e^{-E_n^{\rm GS}(\tau-\tau_*)}\,O(\tau^{-2}) \,.
\end{equation}

Collecting Eqs.~(\ref{eq:zeroth}, \ref{eq:kappat}, \ref{eq:Qresult}), the relative propagator takes the form Eq.~\eqref{eq:GSproj} of the main text,
\begin{equation} \label{eq:knresult}
    k_n(\tyvec,\tau) = \kappa_n(\tau)\, e^{-\tau E_n^{\rm GS}}\left[\psi_n(\tyvec) + O(\tau^{-2})\right] \,,
\end{equation}
where $\kappa_n(\tau)$ includes the constant renormalization of Eq.~\eqref{eq:kappat}. The weakening trap only renormalizes the amplitude, through a time-dependent factor that multiplies the flat and the droplet term of Eq.~\eqref{eq:momentsmapped} alike and therefore drops out of the $n$-derivative in Eq.~\eqref{eq:Sdiffder}, and it leaves the shape unchanged up to $O(\tau^{-2})$.

\textbf{Form of the $O(\tau^{-2})$ correction.} --- The estimate Eq.~\eqref{eq:Qbound} also fixes the shape deformation explicitly. Since only scatterings within $\Delta_{\rm gap}^{-1}$ of $\tau$ contribute, in $\Pi^\perp\ket{\check\Psi^{(1)}(\tau)}$ the incoming state may be replaced by its ground-state part $\c_0e^{-E_n^{\rm GS}(\tau'-\tau_*)}\ket{\Psi_n^{\rm GS}}$ (the rest being smaller by $e^{-\Delta_{\rm gap}(\tau-\tau_*)}$). We can also replace $A(\tau')$ by $A(\tau)$, and the integral over the time $s=\tau-\tau'$ elapsed after the scattering may be extended to infinity, which produces the reduced resolvent of $\hat H_{\rm LL}(1)$. Relative to its ground-state component,
\begin{equation} \label{eq:adiabatic}
\begin{split}
    &\ket{\check\Psi(\tau)} \ \propto\ \ket{\Psi_n^{\rm GS}} \ -\ \frac{A(\tau)}{2}\ \ket{\phi_n}
    \ +\ O\!\left(\frac{A(\tau)}{\Delta_{\rm gap}\,\tau}\right) \ +\ O(A^2) \,, \\
    &\ket{\phi_n} := \frac{\Pi^\perp}{\hat H_{\rm LL}(1)-E_n^{\rm GS}}\,\f\ket{\Psi_n^{\rm GS}} = \sum_{m\neq{\rm GS}}\frac{\bra{m}\f\ket{\Psi_n^{\rm GS}}}{E_n^m-E_n^{\rm GS}}\ket{m} \,,
\end{split}
\end{equation}
where $\{\ket m,E_n^m\}$ are the excited eigenstates and eigenvalues. This is the \emph{static} first-order correction to $\ket{\Psi_n^{\rm GS}}$ produced by the instantaneous perturbation $\tfrac{A(\tau)}{2}\f$: the relative propagator adiabatically follows the ground state of the instantaneous trapped Hamiltonian, and since $A(\tau)\propto(\tau-\tau_0)^{-2}$ the trap produces no $O(\tau^{-1})$ distortion of the shape. The correction is $A(\tau)/2$ times the fixed state $\ket{\phi_n}$, of norm at most $\|\f\ket{\Psi_n^{\rm GS}}\|/\Delta_{\rm gap}(n)$, which corresponds to the $O(\tau^{-2})$ term in Eq.~\eqref{eq:knresult}. Finally, we mention that the first correction term in Eq.~\eqref{eq:adiabatic} is controlled by the adiabatic parameter $(\partial_\tau A)/(A\Delta_{\rm gap})\simeq2/[\Delta_{\rm gap}(\tau-\tau_0)]$, which dominates the second-order Dyson term, $O(A^2)=O(\tau^{-4})$.

\textbf{Replica limit.} --- 
We remark that 
$\Delta_{\rm gap}(n)=n(n-1)/4\to0$ as $n\to1$.
Nonetheless, our construction remains valid as we consider the large-$\tau$ asymptotic at fixed $n$ and only afterwards consider the replica limit $n \to 1$. 
As a matter of fact, the explicit form of the entropy up to the order $O(\tau^{-1})$, retained in the main text, only involves  $\lim_{n\to1} \kappa_n(\tau)$, which trivializes by normalization; the $O(\tau^{-1})$ term of Eq.~\eqref{eq:alphaLT} comes instead from the Gaussian factor $e^{c'\f/(4c)}$ of Eq.~\eqref{eq:momentsmapped} expanded against $\psi_n$, i.e., from the $H^{(2)}_{n-1}$ term of Eq.~\eqref{eq:H2}. Pushing the entropy to $O(\tau^{-2})$ would instead require the analytic continuation in $n$ of the state in Eq.~\eqref{eq:adiabatic}, but would only capture non-universal terms.

\section{The marginal case $\alpha=1/2$} \label{sec:marginal}

At $\alpha=1/2$ the dimensionless coupling $g(t)=c(t)\sqrt t$ of Eq.~\eqref{eq:gdimless} neither flows to zero nor diverges: it sticks to the constant value $g_*= c_0\sqrt{t_0}$, and neither of the two expansions above is available. The mapping of Eq.~\eqref{eq:mapping}, however, still simplifies the problem considerably. Setting $\alpha=1/2$ in Eq.~\eqref{eq:thetadef}, the mapped time grows logarithmically,
\begin{equation}
    \tau(t) = \int_0^{t_0}\!\! ds\ c(s)^2 + c_0^2t_0 \ln \frac{t}{t_0} \ \stackrel{t\to\infty}{\longrightarrow}\ \infty \,,
\end{equation}
while both the trap amplitude and the exponent of the Gaussian factor in Eq.~\eqref{eq:momentsmapped} become time-independent and depend on $g_*$ only,
\begin{equation}
    A = \frac{1}{8 g_*^4} \,,\qquad \frac{c'(\tau)}{4c(\tau)} = -\frac{1}{8g_*^2} = -\sqrt{\frac{A}{8}} \,.
\end{equation}
Since $\tau\to\infty$, the relative propagator is still projected onto the ground state of the LL model, but now in a \emph{static} harmonic trap of amplitude $A$; call $\psi^{(g_*)}_n(\tyvec)$ the corresponding spectral weight, the analogue of Eq.~\eqref{eq:psin}. Repeating the steps of the End Matter, and using $-\ln c(t) = \frac12 \ln t -\ln g_*$, one gets
\begin{equation}
    \s[t] = \frac12 \ln t + \Phi(g_*)\,,\quad
    \Phi(g_*) = -\ln g_* + \partial_n\left[ \int_{\mathbb R^{n-1}}\!\!\! d^{n-1}\tilde y\ e^{-\f(\tyvec)/(8g_*^2)}\psi^{(g_*)}_n(\tyvec) - \psi^{(g_*)}_n(0)\right]_{n=1} \,,
\end{equation}
where the amplitude of the propagator cancels as in Eq.~\eqref{eq:entroexplicit} because the two terms coincide at $n=1$. Hence the leading behavior $\frac12\ln t$ is the same as for $\alpha>1/2$, but the constant is a nonuniversal function of the marginal coupling $g_*$ alone. This is the only point of the phase diagram where the amplitude of the protocol survives in the $O(1)$ term, and it explains why $\Phi(g_*)$ need not match the $\alpha\to1/2^\pm$ limits of Eqs.~(\ref{eq:entroalphalarge},~\ref{eq:alphaLT}), which are attained only beyond the crossover time, itself diverging as $\alpha\to1/2$. Computing $\Phi(g_*)$ requires the $n$-dependence of the trapped LL ground state, which is not known in closed form; the two limits $\Phi(g_*\to0)=\frac12(\ln4\pi+1)+O(g_*)$ and $\Phi(g_*\to\infty)=S_\infty-\ln g_*+O(g_*^{-2})$, matching Eqs.~(\ref{eq:entroalphalarge},~\ref{eq:alphaLT}) at $\alpha\to1/2^\pm$, are however fixed by the two regimes studied above.

\section{Numerics} \label{sec:numerics}

\begin{figure}
    \centering
\includegraphics[width=0.7\linewidth]{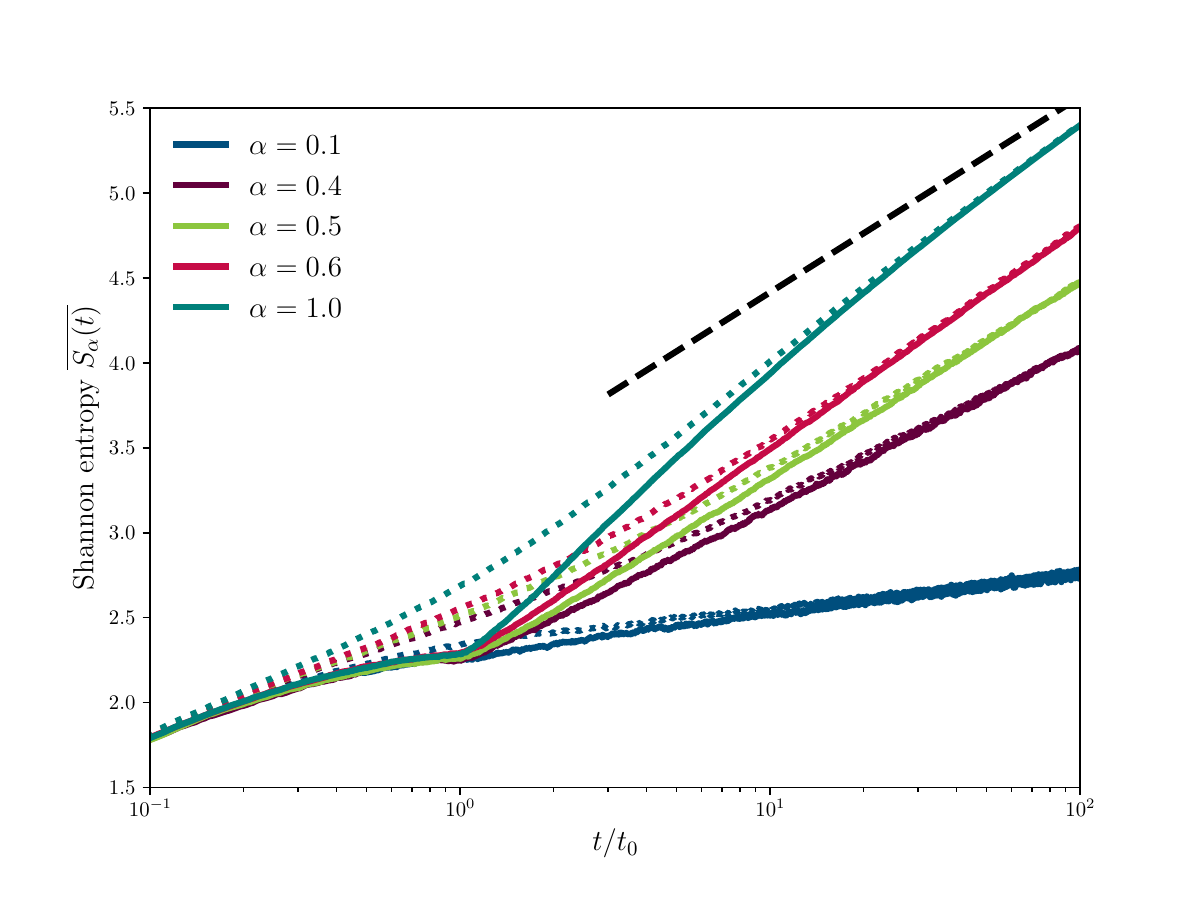}
    \caption{Universality of the long-time behavior for the decreasing-$c(t)$ protocol. Numerical curves ($c_0=1$, $t_0=20$, $\Delta t=5\times 10^{-3}$) for various $\alpha$'s. Solid lines show the protocol where $c(t)=c_0$ before $t_0$, and $c(t)=c_0(t/t_0)^{-\alpha}$ after $t_0$. Dotted lines show $c(t) = c_0(1+t/t_0)^{-\alpha}$.}
    \label{fig:univ}
\end{figure}

In this Section we describe the numerical protocol used to test our analytical results. We realize simulations of the exact discrete-time dynamics described in the main text, where a particle performs a random walk on the one-dimensional lattice, defining a trajectory $(x_\t')_{\t'=0}^\t$. We simulate the particle by setting its initial position $x_0=0$ and the initial occupation probability $p_j^{(0)}=\delta_{j,0}$. Then, we update the particle position by the rule $x_{\t+1} = x_\t\pm 1$ with probability $1/2$ in each case, and evolve the prior distribution $p_j^{(\t+1,-)} = (p_{j-1}^{(\t)}+p_{j+1}^{(\t)})/2$. 
We then perform the position measurements, extracting a set of outcomes $\{a_j^{(\t+1)}\}_{j=-\t}^\t$ from the distributions $P_{s,\t}$, using $P_{1,\t}$ on the occupied site and $P_{0,\t}$ otherwise. 
For the measurement procedure, we consider two paradigmatic examples for the choice of distributions, which provide complementary access to the continuous-time limit and to the universal long-time regime:
\begin{itemize}
\item Choice $(i)$: $P_{s,\t}$ is Gaussian of unit variance and mean $s\beta_\t$:
\begin{equation}
    P_{s,\t}(a) = \frac{1}{\sqrt{2\pi}} \ e^{-\frac{1}{2}(a-s\beta_\t)^2}
    \,,\quad s\in \{0,1\}  \quad \Rightarrow \quad D_{\rm KL}(P_{1,\t}||P_{0,\t}) = \frac{\beta_\t^2}{2} \,,
\end{equation}
In this case, the weak-noise scaling Eq.~\eqref{eq:scaling} in the End Matter, required to define the continuous-time limit, yields $\beta_\t = (\Delta t/2)^{1/4}\sqrt{c(t)}$. Therefore, in the limit of $\Delta t \to 0$, the two distributions overlap completely; 
\item Choice $(ii)$: $P_{s,\t}$ is an inverse-Gamma distribution of parameter $\gamma_\t-s$, i.e., 
\begin{equation}
    P_{s,\t}(a) = \frac{e^{-1/ a}}{ \Gamma(\gamma_\t-s)} a^{-1-(\gamma_\t-s)}
    \,,\quad s\in \{0,1\}  \quad \Rightarrow \quad D_{\rm KL}(P_{1,\t}||P_{0,\t}) =\ln(\gamma_\t-1) - \psi(\gamma_\t-1)  \,,
\end{equation}  
with $\psi$ the digamma function. In the continuous-time limit, we scale the distributions $P_{s,\t}$ with the discrete time-step $\Delta t$ as $\gamma_\t=1+\sqrt{2/\Delta t}/c(t)$. In the limit $\Delta t\to 0$, the two distributions are compressed onto the small values of $a$ and it becomes increasingly harder to distinguish $P_{1,\t}$ from $P_{0,\t}$. In the context of the directed-polymer model, this is usually referred to as the log-gamma polymer.
\end{itemize}
The outcomes are used to compute the posterior distribution, using $p_j^{(\t+1)} \propto P_{1,\t}(a_j^{(\t+1)})/P_{0,\t}(a_j^{(\t+1)}) \ p_j^{(\t+1,-)}$, followed by normalization. 
Notice that, using the Bayesian update Eq.~\eqref{eq:bayesdiscrete}, one could also discard the evolution of the physical particle and extract the $\{a_j^{(\t+1)}\}_{j=-\t}^\t$ from the knowledge of the prior distribution $\{p_j^{(\t+1,-)}\}_{j=-\t}^\t$.
At each time-step, we compute the Shannon entropy Eq.~\eqref{eq:entropydef} using the posterior distribution $S_\t = -\sum_{j=-\t}^\t p_j^{(\t)} \ln p_j^{(\t)}$. We then average over many trajectories to compute the sample mean of the Shannon entropy $\average{S_\t}$.

We choose different values of the time-step, i.e., $\Delta t = 5\times 10^{-2},5\times 10^{-3},5\times 10^{-4}$, showing convergence for both models as $\Delta t$ is decreased. In particular, the convergence to the theoretical results is faster for the Gaussian model (see Fig.~\ref{fig:numerics} in the main text). For the constant-monitoring case, we set $c(t)=\bar c=1$, reproducing the saturation to the universal constant $S_\infty = 1+\gamma_{\rm E} + \ln 2$ once the diverging constant $-\ln 2\Delta x = -\ln 2\sqrt{2\Delta t}$ is removed. The approach to the asymptotics $\delta \mathcal S(t)$ Eq.~\eqref{eq:approach} is harder to test because of the noise on the sampled data, but visible once $\Delta t$ is taken small enough, cf. Fig.~\ref{fig:numerics} (left panel). 

In the case of decreasing $c(t)$ with power-law tail (Eq.~\eqref{eq:ct} in the main text), we test various protocols. We show agreement between the Gaussian and log-gamma models, and convergence to the theoretical results. In order to test universality, we also choose different early-time shapes for $c(t)$. In particular, we consider the cases where $c(t) = c_0$ before $t_0$ (this case is used to produce the plots shown in Fig.~\ref{fig:numerics} in the main text) and where $c(t) = c_0(1+t/t_0)^{-\alpha}$ for all times. Both protocols share the same power-law tail and differ only at early times, and they converge to the same asymptotic regimes, see Fig.~\ref{fig:univ}: this is a direct numerical test of the universality statement of the main text, namely that the late-time entropy depends on the protocol only through the combination $c_0t_0^\alpha$.
Moreover, we test our results for various values of $\alpha$ and $t_0$. 

\end{document}